\documentclass[showpacs,preprintnumbers,amsmath,amssymb,nofootinbib,floatfix]{revtex4}

\usepackage[dvips]{graphicx}
\usepackage{dcolumn}
\usepackage{bm}

\def\mapgeq{\mathbin{\lower.3ex\hbox{$\buildrel>\over{\smash{\scriptstyle\sim}\vphantom{_x}}$}}}
\def\mapleq{\mathbin{\lower.3ex\hbox{$\buildrel<\over{\smash{\scriptstyle\sim}\vphantom{_x}}$}}}
\def\mapgeqeq{\mathbi{\lower.3ex\hbox{$\buildrel>\over{\smash{\scriptstyle\approx}\vphantom{_2}}$}}}
\def\mapleqeq{\mathbin{\lower.3ex\hbox{$\buildrel<\over{\smash{\scriptstyle\approx}\vphantom{_2}}$}}}
\def\Journal#1#2#3#4{{#1} {\bf #2}, #3 (#4)}
\def\MPL{Mod. Phys. Lett. A}

\def\NPB{Nucl. Phys. B}
\def\NPBOLD{Nucl. Phys.}

\def\PLB{{Phys. Lett.} B}

\def\PLBOLD{Phys. Lett.}
\def\PRL{Phys. Rev. Lett.}
\def\RMP{Rev. Mod. Phys.}
\def\PRD{Phys. Rev. D}

\def\PTP{Prog. Theor. Phys.}
\def\JHEP{JHEP}
\def\JCAP{JCAP}
\def\EPJ{Euro. Phys. J. C}
\def\EPJA{Euro. Phys. J. A}
\def\JETPUSSR{Sov. Phys. JETP}
\def\ZETP{Zh. Eksp. Teor. Piz.}

\def\IJMP{Int. J. Mod. Phys. A}

\def\JPG{J. Phys. G}
\def\PAN{Phys. At. Nucl.}
\def\SCI{Science}
\def\APJ{Astrophysics J.}

\def\Erratum{Erratum-ibid}

\begin{document}

\preprint{TOKAI-HEP/TH-0501}

\title{Constraints on Flavor Neutrino Masses and $\sin^22\vartheta_{12}\gg\sin^2\vartheta_{13}$ in Neutrino Oscillations}

\author{Ichiro Aizawa}
 \email{4aspd001@keyaki.cc.u-tokai.ac.jp}

\author{Teruyuki Kitabayashi$^a$\footnote{Address after April 1, 2005: Department of Physics, Tokai University, 1117 Kitakaname, Hiratsuka, Kanagawa 259-1291, Japan}}
 \email{teruyuki@keyaki.cc.u-tokai.ac.jp}

\author{Masaki Yasu\`{e}}%
\email{yasue@keyaki.cc.u-tokai.ac.jp}
\affiliation{\vspace{5mm}%
\sl Department of Physics, Tokai University,\\
1117 Kitakaname, Hiratsuka, Kanagawa 259-1291, Japan\\
\\
$^a$\sl Accelerator Engineering Center \\
Mitsubishi Electric System \& Service Engineering Co.Ltd.\\
2-8-8 Umezono, Tsukuba, Ibaraki 305-0045, Japan}

\date{February, 2005}

\begin{abstract}
To realize the condition of $\sin^22\vartheta_{12}\gg\sin^2\vartheta_{13}$, we find constraints on flavor neutrino masses $M_{ij}$ ($ij=e,\mu,\tau$): C1) $c_{23}^2 M_{\mu\mu} + s_{23}^2 M_{\tau\tau} \approx 2 s_{23} c_{23}M_{\mu\tau} + M_{ee}$ and/or C2) $\vert c_{23}M_{e\mu} -s_{23}M_{e\tau}\vert \gg \vert s_{23}M_{e\mu} +c_{23}M_{e\tau}\vert$, where $c_{23}=\cos\vartheta_{23}$ ($s_{23}=\sin\vartheta_{23}$) and $\vartheta_{12}$, $\vartheta_{13}$ and $\vartheta_{23}$ are the mixing angles for three flavor neutrinos. The applicability of C1) and C2) is examined in models with one massless neutrino and two massive neutrinos suggested by $\det(M)=0$, where $M$ is a mass matrix constructed from $M_{ij}$ ($i,j=e,\mu,\tau$).  To make definite predictions on neutrino masses and mixings, especially on $\sin\vartheta_{13}$, that enable us to trace C1) and C2), $M$ is assumed to possess texture zeros or to be constrained by textures with $M_{\mu\mu}=M_{\tau\tau}$ or $M_{e\tau}=\pm M_{e\mu}$ which turn out to ensure the emergence of the maximal atmospheric neutrino mixing at $\sin\vartheta_{13}\rightarrow 0$.  It is found that C1) is used by textures such as $M_{e\mu}$=0 or $M_{e\tau}$=0 while C2) is used by textures such as $M_{e\tau}=\pm M_{e\mu}$. 
\end{abstract}

\maketitle
\section{\label{sec:1}Introduction}
Since the first experimental confirmation of atmospheric neutrino oscillations by Super-Kamiokande \cite{SK}, there have been various other neutrino oscillations found in solar neutrinos who have given indications for long time \cite{OldSolar}, accelerator neutrinos and reactor neutrinos \cite{experiments}. Oscillations of the terrestrial neutrinos turn out to be identical to those of the neutrinos created by the Nature. It is the origin of neutrino oscillations that neutrino have masses \cite{PMNS}, which can be created by seesaw mechanism \cite{Seesaw,type2seesaw} or by radiative mechanism \cite{Zee,Babu}. These neutrino oscillations are explained by the mixings among the known three flavor neutrinos, $\nu_{e,\mu,\tau}$, which are finally converted into three massive neutrinos, $\nu_{1,2,3}$, and are described by their masses of $m_{1,2,3}$ and mixing angles of $\vartheta_{12,23,13}$. The current experimental data of the neutrino oscillations are characterized by the square mass differences for atmospheric neutrinos $\Delta m_{atm}^2$ with the mixing angle of $\vartheta_{atm}$ and for the solar neutrinos $\Delta m_\odot^2$ with the mixing angle of $\vartheta_\odot$.  The CHOOZ collaboration has tried to measure another mixing angle $\vartheta_{CHOOZ}$. The result of the observations can be summarized in the following values \cite{NeutrinoSummary}:
\begin{eqnarray}
&&5.4 \times 10^{-5} eV^2 < \Delta m_\odot^2 < 9.5 \times 10^{-5} eV^2, \quad
0.70 < \sin^2 2\vartheta_\odot < 0.95,
\nonumber \\
&&1.2 \times 10^{-3} eV^2 < \Delta m_{atm}^2 < 4.8 \times 10^{-3} eV^2, \quad
0.92 < \sin^22\vartheta_{atm},
\nonumber \\
&& \sin\vartheta_{CHOOZ} <0.23.
\label{Eq:Data}
\end{eqnarray}
These observed $\Delta m^2_{atm,\odot}$ and the mixing angles can be identified with $\Delta m_{atm}^2= \vert m_3^2-m_2^2\vert$, $\Delta m_\odot^2=\vert m_2^2-m_1^2\vert$, $\vartheta_{atm}=\vartheta_{23}$, $\vartheta_\odot=\vartheta_{12}$ and $\vartheta_{CHOOZ}=\vartheta_{13}$.  It should be noted that, from the estimation of effects of matter on solar neutrinos, the mass eigenstate of the larger electron neutrino components $\nu_1$ has the smaller mass than that of the smaller electron neutrino components $\nu_2$ \cite{PositiveSolor}.  Thus, the sign of $m_2^2 - m_1^2$ is positive so that $\Delta m_\odot^2 = m_2^2 - m_1^2$ $(> 0)$.

One of the remarkable features of the observed neutrino oscillations lies in the fact that $\sin^22\vartheta_{12,23} \gg \sin^2\vartheta_{13}$.  The almost maximal atmospheric neutrino mixing of $\sin^22\vartheta_{23}\sim 1$ may arise as a result of the presence of an approximate $\mu$-$\tau$ symmetry \cite{Nishiura,mu-tau,mu-tau1} in flavor neutrino masses forming a symmetric mass matrix of $M$ in the ($\nu_e$, $\nu_\mu$, $\nu_\tau$)-basis. Namely, the requirement of either $M_{e\mu}$=$\pm M_{e\tau}$ or $M_{\mu\mu}$=$M_{\tau\tau}$ ensures the appearance of the maximal atmospheric neutrino mixing at the limit of $\sin\vartheta_{13}\rightarrow 0$, where $M_{ij}$ ($i,j$=$e,\mu,\tau$) is the flavor neutrino mass for $\nu_i\nu_j$ as an $ij$-matrix element of $M$.  However, the reason for the large but not maximal mixing of $\sin^22\vartheta_{12}$ is not well understood.  Some possible answers would be based on the bimaximal mixing scheme \cite{Bimaximal} relying upon the $L_e-L_\mu-L_\tau$ conservation \cite{Lprime} and on the tri-bimaximal mixing scheme \cite{tri-bimaximal,tri-bimaximal-review,NearlyBimaximal}, which, respectively, yield $\sin^22\vartheta_{12}=1$ with $\sin^22\vartheta_{23}$ left undetermined and $\sin^22\vartheta_{12}=8/9$ with $\sin^22\vartheta_{23}=1$.  Before employing such specific schemes, we would like to make more general discussions that guide us to understand the possible form of $M$ to explain the fact of $\sin^22\vartheta_{12}\gg\sin^2\vartheta_{13}$.

In this paper, we present expressions to determine $\vartheta_{12}$ and $\vartheta_{13}$ in terms of $M_{ij}$, which are given by
\begin{eqnarray}
&& \tan2\vartheta_{12} \approx \frac{2\left(c_{23}M_{e\mu}-s_{23}M_{e\tau}\right)}{c_{23}^2 M_{\mu\mu} + s_{23}^2 M_{\tau\tau} - 2 s_{23} c_{23}M_{\mu\tau} - M_{ee}},
\label{Eq:MixingAnglesSol}
\end{eqnarray}
under the approximation of $\sin^2\vartheta_{13}\approx 0$, and 
\begin{eqnarray}
&& \tan2\vartheta_{13} = \frac{2\left(s_{23}M_{e\mu} +c_{23}M_{e\tau}\right)}{s_{23}^2 M_{\mu\mu} + c_{23}^2 M_{\tau\tau} + 2 s_{23} c_{23}M_{\mu\tau} - M_{ee}}.
\label{Eq:MixingAnglesCHOOZ}
\end{eqnarray}
From these expressions, the conditions to have $\vert\tan 2\vartheta_{12}\vert\gg\vert\tan 2\vartheta_{13}\vert$, namely, $\sin^22\vartheta_{12}\gg\sin^2\vartheta_{13}$, can be translated into the conditions among the matrix elements of $M$. Then, we are able to infer some general underlying properties of $M$. To have $\vert\tan 2\vartheta_{12}\vert\gg\vert\tan 2\vartheta_{13}\vert$ can be achieved by the constraints of 
\begin{itemize}
\item C1) $c_{23}^2 M_{\mu\mu} + s_{23}^2 M_{\tau\tau} - 2 s_{23} c_{23}M_{\mu\tau} - M_{ee}\approx 0$ and/or
\item C2) $\vert c_{23}M_{e\mu}-s_{23}M_{e\tau}\vert \gg \vert s_{23}M_{e\mu} +c_{23}M_{e\tau}\vert$.
\end{itemize}
It should be noted that the sign of $s_{23}$ is crucial for C2) to satisfy $\vert c_{23}M_{e\mu}-s_{23}M_{e\tau}\vert \gg \vert s_{23}M_{e\mu} +c_{23}M_{e\tau}\vert$ because the cancellation may occur in $s_{23}M_{e\mu} +c_{23}M_{e\tau}$ to reduce its magnitude if the sign of $s_{23}$ is reversed. To see if the requirements from C1) and C2) are plausible, we use explicit models to calculate $\tan 2\vartheta_{13}$ and compare it with $\tan 2\vartheta_{12}$.  

Since properties of neutrino oscillations have been clarified in detail by continuous efforts of observing neutrino oscillations \cite{RecentAnalyses}, the matrix element of $M$ can be completely determined in principle. However, the neutrino mass matrix $M$ at least has 6 parameters and all of these parameters cannot be fixed at present. To somehow determine the texture of $M$, we need constraints on $M$, which reduce the number of the independent parameters in $M$, so that the texture can be determined by the observed data.  For example, such constraints are supplied by the condition that the texture includes some zeros in matrix elements of $M$ \cite{ZeroTexture0,ZeroTexture} and by the flavor-basis independent conditions on $M$ such as $\det (M)= m_1m_2m_3=0$ \cite{detM} and ${\rm tr} (M)=m_1+m_2+m_3=0$ \cite{trM}.  The condition of $\det (M)=0$ is equivalent to at least demand the presence of one massless neutrino. It is further recognized that even if one of the neutrino mass eigenstates, $m_1$ or $m_3$, is exactly zero, we can still explain the observed neutrino oscillation data \cite{detM}. This fact suggests that the condition of $\det (M)=0$ is not merely an artificial assumption to reduce the number of independent parameters but can be regarded as a physical assumption that requires the presence of one massless neutrino (or one extremely light neutrino) in the Nature.  In fact, the extremely light neutrino of $m_1\sim 10^{-10}$ eV may account \cite{Leptogenesis} for the Affleck-Dine scenario \cite{Affleck-Dine} for leptogenesis \cite{LeptogenesisFirst}.

We would like to examine possible textures consistent with the observed neutrino properties and to predict allowed regions of the phenomenologically viable angle of $\vartheta_{13}$ \cite{theta-13-th,theta-13-ex} to be compared with $\sin^22\vartheta_{12}$. To see the applicability of C1) and C2) in various textures, we directly compute flavor neutrino masses, where we can trace the source generating $\vert\tan 2\vartheta_{12}\vert\gg\vert\tan 2\vartheta_{13}\vert$ to see how C1) and C2) work. We focus on models, where one massless neutrino with either $m_1=0$ or $m_3=0$ is present, and examine textures with one vanishing matrix element and with some relations between matrix elements that ensure the appearance of the maximal atmospheric neutrino mixing in the limit of $\sin\theta_{13}\rightarrow 0$.  The latter textures are suggested by the $\mu$-$\tau$ symmetry \cite{Nishiura,mu-tau,mu-tau1}.  From Eq.(\ref{Eq:MixingAnglesCHOOZ}), we will find that the suppressed magnitude of $\sin\vartheta_{13}$ of ${\mathcal{O}}$(0.1) reflects that of $M_{e\mu,e\tau}$ and that to obtain more suppressed magnitude of $\sin\vartheta_{13}$ of ${\mathcal{O}}$(0.01) needs the cancellation due to $s_{23}M_{e\mu} +c_{23}M_{e\tau}\sim 0$.  For instance, it will be shown that the texture with $M_{e\mu}=0$ (or $M_{e\tau}=0$), which cannot have $s_{23}M_{e\mu} +c_{23}M_{e\tau}\sim 0$, turns out to ``naturally" predict the larger magnitude of $\sin\vartheta_{13}\mapgeq 0.1$ in the so-called normal mass hierarchy ($NMH$) with $\vert m_1\vert (=0) \ll \vert m_2\vert \ll \vert m_3\vert$ while, in the so-called inverted mass hierarchy ($IMH$) with $\vert m_1\vert \sim \vert m_2\vert \gg \vert m_3\vert (=0)$, the observed data is explained by $M_{e\mu}$ or $M_{e\tau}$ itself of ${\mathcal{O}}$(0.01) in the same textures, which is ``unnaturally" suppressed.

In the next section,  we discuss general theoretical consequences of the form of $M$ constrained from the experimental data, which include the useful expression of $\tan 2\vartheta_{12,13}$. In Sec.\ref{sec:3}, we examine each texture to calculate masses and mixing angles, especially to clarify how to yield the suppressed $\sin^2\theta_{13}$ and to find the possible origin for $\sin^22\vartheta_{12}\gg\sin\vartheta_{13}$ in various allowed textures. The final section is devoted to summary and discussions.

\section{\label{sec:2}General Constraints on Flavor Neutrino Masses}
Before going into discussions of specific models, where one massless neutrino is present, we use general formulas to add constraints on the masses and mixing angles in terms of the flavor neutrino masses.  Our flavor neutrino mass matrix, $M$, is defined by\footnote{It is understood that the charged leptons and neutrinos are rotated, if necessary, to give diagonal charged-current interactions and to define $\nu_e$, $\nu_\mu$ and $\nu_\tau$.}
\begin{eqnarray}
&& M = \left( {\begin{array}{*{20}c}
	M_{ee} & M_{e\mu} & M_{e\tau}  \\
	M_{e\mu} & M_{\mu\mu} & M_{\mu\tau}  \\
	M_{e\tau} & M_{\mu\tau} & M_{\tau\tau}  \\
\end{array}} \right).
\label{Eq:NuMatrixEntries}
\end{eqnarray}
The flavor neutrinos $\vert \nu_{\ell=e,\mu,\tau} \rangle$ and their mass eigenstates $\vert \nu_{i=1,2,3} \rangle$ are known to be related by $\vert \nu_\ell \rangle = U_{\ell i} \vert \nu_i \rangle$ to yield $U^TMU$=diag.$(m_1,m_2,m_3)$ \cite{PMNS}, where $U$ represents the unitary transformation usually parameterized by
\begin{eqnarray}
U &=&
    \left( 
    \begin{array}{ccc}
    c_{12}c_{13}                     &  s_{12}c_{13}   &  s_{13}\\
    -s_{12}c_{23}-c_{12}s_{23}s_{13} &  c_{12}c_{23}-s_{12}s_{23}s_{13}  &  s_{23}c_{13}\\
    s_{12}s_{23}-c_{12}c_{23}s_{13}  &  -c_{12}s_{23}-s_{12}c_{23}s_{13} & c_{23}c_{13}\\
    \end{array} 
    \right),
\end{eqnarray}
where $c_{ij} \equiv \cos\vartheta_{ij}$ and $s_{ij} \equiv \sin\vartheta_{ij}$, and we assume no CP violation in the lepton sector.\footnote{For $M$ allowing CP-violation, see discussions in Ref.\cite{GeneralM-CP}.}  

From the results in the Appendix \ref{sec:Appendix}, we can derive the following important properties of $M$. In the ideal case with $\sin\vartheta_{13}$=0, Eq.(\ref{Eq:MixingAngles}) demands that $s_{23}M_{e\mu} +c_{23}M_{e\tau}$=0, from which the maximal atmospheric mixing with $c_{23} = \sigma s_{23} = 1/\sqrt{2}$ for $\sigma=\pm 1$ can be derived by requiring that $M_{e\tau}= -\sigma M_{e\mu}$.  Furthermore, Eq.(\ref{Eq:f-d}) with the maximal atmospheric mixing reads $M_{\tau\tau} = M_{\mu\mu} + 2s_{13}X$ (unless $M_{\mu\tau}=0$), from which the requirement of $M_{\tau\tau} = M_{\mu\mu}$ yields either $X(=c_{23}M_{e\mu} -s_{23}M_{e\tau})=0$ leading to $M_{e\mu} =\sigma M_{e\tau}$ or $\sin\vartheta_{13}$=0.\footnote{For the texture with $M_{\mu\mu~{\rm or}~\tau\tau}=0$, the maximal atmospheric neutrino mixing cannot be realized as a result of $\cos 2\vartheta_{23}= 0$, which does not satisfy Eq.(\ref{Eq:f-d}). Instead, $\sigma M_{e\mu} +M_{e\tau}\sim 0$ is to be satisfied. It is understood that this texture is excluded for $M_{\mu\mu}\sim M_{\tau\tau}$.}  Therefore, we observe that the almost maximal atmospheric mixing can arise if
\begin{eqnarray}
	M_{e\tau}\approx -\sigma M_{e\mu}~{\rm and/or}~M_{\tau\tau} \approx M_{\mu\mu},
\label{Eq:Conditions1}
\end{eqnarray}
as $\sin\vartheta_{13}\rightarrow 0$ and if
\begin{eqnarray}
	M_{e\tau}\approx \sigma M_{e\mu}~{\rm and}~M_{\tau\tau} \approx M_{\mu\mu},
\label{Eq:Conditions2}
\end{eqnarray}
irrespective of the magnitude of $\sin\vartheta_{13}$.  Because $X\approx 0$ in Eq.(\ref{Eq:Conditions2}), giving $m_1\approx m_2$ from Eq.(\ref{Eq:TwoMasses1-2}), the condition of Eq.(\ref{Eq:Conditions2}) is only possible to be satisfied in $IMH$ with $\vert m_1\vert \sim\vert m_2\vert \gg\vert m_3\vert$.  If $M_{\tau\tau} \approx M_{\mu\mu}$ and $\vert M_{e\mu}\vert \approx \vert M_{e\tau}\vert$ are simultaneously satisfied, these relations imply the presence of the (approximate) $\mu$-$\tau$ permutation symmetry for the neutrino masses \cite{Nishiura,mu-tau,mu-tau1}.  This realization of $\sin^22\vartheta_{23}\sim 1$ and $\sin\vartheta_{13}\sim 0$ is in a sense ``natural" because it arises from the consequence of the symmetry principle.  However, it may be a real physics that accidentally yields $M_{\mu\mu}\sim M_{\tau\tau}$ for $\sin^22\vartheta_{23}\sim 1$ with a suppressed magnitude of $\sin\vartheta_{13}$, which demands the suppression of $M_{e\mu,e\tau}$ itself:
\begin{eqnarray}
	&& \vert s_{23}^2 M_{\mu\mu} + c_{23}^2 M_{\tau\tau} + 2 s_{23} c_{23}M_{\mu\tau} - M_{ee}\vert \gg \vert M_{e\mu}\vert , \vert M_{e\tau} \vert,
\label{Eq:Small-13-Condition}
\end{eqnarray}
in $\tan 2\vartheta_{13}$ of Eq.(\ref{Eq:MixingAngles}). A typical example to be found is a texture with $M_{e\mu}=0$ or $M_{e\tau}=0$ for $IMH$. 

In $NMH$, since $\Delta m^2_{atm}\gg\Delta m^2_\odot$, we find that $\vert m_{1,2}\vert \ll \vert m_3\vert$, which becomes $\vert \lambda_{1,2}\vert \ll \vert \lambda_3\vert$ because of Eq.(\ref{Eq:TwoMasses1-2Simplified}) and $m_3 \sim \lambda_3$ for $\sin^2\vartheta_{13}\ll 1$. The requirement of $\vert \lambda_2\vert \ll \vert \lambda_3\vert$ gives $\vert c_{23}^2 M_{\mu\mu} + s_{23}^2 M_{\tau\tau} - 2 s_{23} c_{23}M_{\mu\tau}\vert \ll \vert s_{23}^2 M_{\mu\mu} + c_{23}^2 M_{\tau\tau} + 2 s_{23} c_{23}M_{\mu\tau}\vert$, which roughly yields,
\begin{eqnarray}
&&c_{23}^2 M_{\mu\mu} + s_{23}^2 M_{\tau\tau} \sim 2 s_{23} c_{23}M_{\mu\tau},
\label{Eq:NMH-Condition}
\end{eqnarray}
leading to
\begin{eqnarray}
&&M_{\mu\mu} + M_{\tau\tau} \sim 2 \sigma M_{\mu\tau},
\label{Eq:NMH-Condition-AlmostMaximal}
\end{eqnarray}
for $c_{23}\sim \sigma s_{23}\sim1/\sqrt{2}$.  It is satisfied by textures of $NMH$ to be discussed in Sec.\ref{sec:3}.  Since $M_{\mu\mu}\sim M_{\tau\tau}$ from Eq.(\ref{Eq:f-d}) for $c_{12}\sim \sigma s_{23} \sim 1/\sqrt{2}$ (except for $M_{\mu\mu~{\rm or}~\tau\tau}=0$), it implies a more constrained relation of $M_{\mu\mu} \sim M_{\tau\tau} \sim \sigma M_{\mu\tau}$ or even the ideal situation of
\begin{eqnarray}
&& M_{\mu\mu} = M_{\tau\tau} = \sigma M_{\mu\tau},
\label{Eq:IdealNMHMij}
\end{eqnarray}
that suggests the presence of democratic interactions among $\nu_{\mu,\tau}$ in neutrino physics. 

Similarly, in $IMH$, from $\Delta m^2_{atm}\gg\Delta m^2_\odot$, we find that $\vert m_1\vert \sim \vert m_2\vert \gg \vert m_3\vert$ requiring $m_1\sim \pm m_2$, which becomes $\vert \lambda_{1,2}\vert \gg \vert \lambda_3\vert$.  The requirement of $\vert \lambda_{1,2}\vert \gg \vert \lambda_3\vert$ instead gives $\lambda_3\sim 0$
\begin{eqnarray}
&&s_{23}^2 M_{\mu\mu} + c_{23}^2 M_{\tau\tau} \sim -2 s_{23} c_{23}M_{\mu\tau},
\label{Eq:IMH-Condition}
\end{eqnarray}
leading to
\begin{eqnarray}
&&M_{\mu\mu} + M_{\tau\tau} \sim -2 \sigma M_{\mu\tau}.
\label{Eq:IMH-Condition-AlmostMaximal}
\end{eqnarray}
The textures with $IMH$ are to be found to generally exhibit Eq.(\ref{Eq:IMH-Condition-AlmostMaximal}). Again, the situation of 
\begin{eqnarray}
&& M_{\mu\mu} = M_{\tau\tau} = -\sigma M_{\mu\tau}
\label{Eq:IdealIMHMij}
\end{eqnarray}
may be realized in ``ideal" neutrino physics.  Furthermore, Eq.(\ref{Eq:TwoMasses1-2Simplified}) leads to $\lambda_1\sim \lambda_2$ for $m_1\sim m_2$ provided that $\vert\lambda_1+\lambda_2\vert \gg \vert X\vert$ and to $\lambda_1\sim -\lambda_2$ for $m_1\sim -m_2$ provided that $\vert\lambda_1+\lambda_2\vert \ll \vert X\vert$.  In terms of $M_{ij}$, these constraints can be expressed as
\begin{eqnarray}
&& c^2_{23}M_{\mu\mu}+s^2_{23}M_{\tau\tau}-2s_{23}c_{23} M_{\mu\tau} \sim \pm M_{ee},
\label{Eq:IMH-Condition-2}
\end{eqnarray}
for $m_1\sim \pm m_2$.  For $c_{23}\sim \sigma s_{23}\sim1/\sqrt{2}$, it leads to
\begin{eqnarray}
&& M_{\mu\mu}+M_{\tau\tau}-2\sigma M_{\mu\tau} \sim \pm 2M_{ee}.
\label{Eq:IMH-Condition-AlmostMaximal-2}
\end{eqnarray}
An additional constraint on $\lambda_{1,2}$ arises in the case of $\vert X\vert \ll \vert \lambda_1+\lambda_2\vert$ ($m_1\sim m_2$) suggesting $X\sim 0$.  We will find that, in the textures such as those with $M_{e\mu}=0$ and $M_{e\tau}=0$ giving $X$=$c_{23}M_{e\mu}/c_{13}$ and $-s_{23}M_{e\tau}/c_{13}$, the large suppression of $X$ directly requires the large suppression of $M_{e\mu}$ or $M_{e\tau}$ itself, which is the similar to Eq.(\ref{Eq:Small-13-Condition}).  While if $M_{e\mu}/M_{e\tau} \sim \tan \vartheta_{23}$ can be satisfied, the large suppression is not required to give $X\sim 0$ such as in the textures with $M_{e\tau}=\pm M_{e\mu}$.

Once these results are obtained, it is useful to express the flavor neutrino masses as follows:
\begin{eqnarray}
&& M_{ee}  = c_{13}^2 \left( {c_{12}^2 m_1  + s_{12}^2 m_2 } \right) + s_{13}^2 m_3,
\nonumber \\ 
&& M_{e\mu }  = s_{23} s_{13}c_{13} \left[ {m_3  - \left( {c_{12}^2 m_1  + s_{12}^2 m_2 } \right)} \right] -  s_{12}c_{12} c_{23} c_{13} \left( m_1  - m_2 \right),
\nonumber \\ 
&& M_{e\tau }  = c_{23} s_{13} c_{13} \left[ {m_3  - \left( {c_{12}^2 m_1  + s_{12}^2 m_2 } \right)} \right] + s_{12} c_{12} s_{23} c_{13} \left( m_1  - m_2 \right),
\nonumber \\ 
&& M_{\mu \mu }  = c_{23}^2 \left( {s_{12}^2 m_1  + c_{12}^2 m_2 } \right) + s_{23}^2 s_{13}^2 \left( {c_{12}^2 m_1  + s_{12}^2 m_2 } \right) + s_{23}^2 c_{13}^2 m_3  + 2s_{12}c_{12}  s_{23}c_{23}  s_{13} \left( m_1  - m_2 \right),
\nonumber \\ 
&& M_{\mu \tau }  =  s_{23}c_{23} \left[ {s_{13}^2 \left( {c_{12}^2 m_1  + s_{12}^2 m_2 } \right) - \left( {s_{12}^2 m_1  + c_{12}^2 m_2 } \right) + c_{13}^2 m_3 } \right] + s_{12}c_{12}  \left( {c_{23}^2  - s_{23}^2 } \right)s_{13} \left( {m_1  - m_2 } \right),
\nonumber \\ 
&& M_{\tau \tau }  = s_{23}^2 \left( {s_{12}^2 m_1  + c_{12}^2 m_2 } \right) + c_{23}^2 s_{13}^2 \left( {c_{12}^2 m_1  + s_{12}^2 m_2 } \right) + c_{23}^2 c_{13}^2 m_3  - 2 s_{12}c_{12}   s_{23}c_{23} s_{13} \left( m_1  - m_2 \right).
\label{Eq:MassMatrixElements}
\end{eqnarray}
Using $c_{23}\sim\sigma s_{23}\sim 1/\sqrt{2}$ and $s_{13}\sim 0$ as well as $\vert m_1\vert < \vert m_2\vert\ll \vert m_3 \vert$ in $NMH$ and $\vert m_2 \vert > \vert m_1\vert \gg \vert m_3\vert$ in $IMH$, we can readily understand that why the obtained constraints are so satisfied.

In addition to these constraints, from Eq.(\ref{Eq:MixingAngles}), we notice another constraint derived from the similarity between the expressions of $\tan 2\vartheta_{12}$ and $\tan 2\vartheta_{13}$, which are given by Eq.(\ref{Eq:MixingAnglesSol}) because of $\lambda_1\sim M_{ee}$ for $\sin^2\vartheta_{13}\sim 0$ and Eq.(\ref{Eq:MixingAnglesCHOOZ}), which turn out to be more symmetric expressions:
\begin{eqnarray}
&& \tan2\vartheta_{12} \sim \frac{2\sqrt{2}\left(M_{e\mu}-\sigma M_{e\tau}\right)}{M_{\mu\mu} + M_{\tau\tau} - 2 \sigma M_{\mu\tau} - 2M_{ee}},
\label{Eq:MixingAngles12-2} \\
&& \tan2\vartheta_{13} \sim \frac{2\sqrt{2}\left(\sigma M_{e\mu} +M_{e\tau}\right)}{ M_{\mu\mu} + M_{\tau\tau} + 2 \sigma M_{\mu\tau} - 2M_{ee}},
\label{Eq:MixingAngles13-2}
\end{eqnarray}
under the approximation of $c_{23}\sim \sigma s_{23}\sim 1/\sqrt{2}$.  This similarity is useful to see how $\sin^22\vartheta_{12}\gg \sin^2\vartheta_{13}$ is realized.  As stated in the Introduction, the relation of $\sin^22\vartheta_{12}\gg \sin^2\vartheta_{13}$ is typically expected to arise from the constraints of C1) and C2), where $s_{23}M_{e\mu} +c_{23}M_{e\tau}\sim 0$ is probable for the latter case. In $IMH$, combined with Eq.(\ref{Eq:IMH-Condition-2}), these conditions lead to C1) for $m_1\sim m_2$ and C2) with $s_{23}^2 M_{\mu\mu} + c_{23}^2 M_{\tau\tau} \approx 2 s_{23} c_{23}M_{\mu\tau} - M_{ee}$ for $m_1\sim -m_2$.  If C1) is realized, we obtain that $\lambda_3\sim M_{\mu\mu} +  M_{\tau\tau}  - M_{ee}$ and
\begin{eqnarray}
\tan2\vartheta_{13} \approx \frac{2\left(s_{23}M_{e\mu} +c_{23}M_{e\tau}\right)}{ M_{\mu\mu} +  M_{\tau\tau}  - 2 M_{ee}}.
\label{Eq:MixingAngles13-0}
\end{eqnarray}
In $IMH$, since $\lambda_1\sim M_{ee}$ for $\sin^2\vartheta_{13}\sim 0$, $\lambda_3\sim m_3$ and $(\vert m_3\vert\ll) \vert m_{1,2}\vert\sim\vert\lambda_{1,2}\vert$ from Eq.(\ref{Eq:TwoMasses1-2Simplified}), we obtain $\vert \lambda_3\vert\ll\vert M_{ee}\vert$, which directly gives
\begin{eqnarray}
\tan2\vartheta_{13} \approx -\frac{2\left(s_{23}M_{e\mu} +c_{23}M_{e\tau}\right)}{M_{ee}},
\label{Eq:MixingAngles13IMH}
\end{eqnarray}
from $\tan 2\vartheta_{13}$ of Eq.(\ref{Eq:MixingAngles}).

We have sufficient conditions that can be reshuffled to give other useful constraints. In the case of $M_{\mu\mu}\sim M_{\tau\tau}$ for the almost maximal atmospheric neutrino mixing, two conditions of $M_{\mu\mu} + M_{\tau\tau} \sim 2 \sigma M_{\mu\tau} + 2M_{ee}$ for $\sin^22\vartheta_{12}\gg \sin^2\vartheta_{13}$ as C1) and $M_{\mu\mu} + M_{\tau\tau} \sim 2 \sigma M_{\mu\tau}$ for $\vert\Delta m^2_{atm}\vert \gg\Delta m^2_\odot$ yield
\begin{eqnarray}
\vert M_{\mu\mu}+M_{\tau\tau}\vert \gg \vert M_{ee}\vert,
\label{Eq:GeneralNMH}
\end{eqnarray}
in $NMH$.  On the other hand, in $IMH$, those of $M_{\mu\mu} + M_{\tau\tau} \sim 2 \sigma M_{\mu\tau} \pm 2M_{ee}$ for $m_1\sim \pm m_2$ and $M_{\mu\mu} + M_{\tau\tau} \sim -2 \sigma M_{\mu\tau}$ for $\vert\Delta m^2_{atm}\vert \gg\Delta m^2_\odot$ yield
\begin{eqnarray}
M_{\mu\mu}+M_{\tau\tau} \sim \pm M_{ee}, \quad 2\sigma M_{\mu\tau} \sim \mp M_{ee}.
\label{Eq:GeneralIMH}
\end{eqnarray}
The texture with $M_{ee}=0$ in $IMH$ does not explain $\vert m_1\vert \sim \vert m_2\vert$. For $m_1\sim m_2$, owing to the constraint of $M_{\mu\mu} +  M_{\tau\tau}  \sim M_{ee}$, Eqs.(\ref{Eq:MixingAngles13-0}) and (\ref{Eq:MixingAngles13IMH}) applied to $IMH$ become consistent with each other.  These results are summarized in TABLE \ref{Tab:Masses}. 
 
To see how $\sin^22\vartheta_{12}\gg \sin^2\vartheta_{13}$ is numerically obtained, we use the ``theoretical" data to be collected in the next section, where we assume the interesting possibility that the Nature enjoys the presence of one massless neutrino. For the rest of discussions, we employ models with texture zeros and with $M_{\mu\mu}$=$M_{\tau\tau}$ and $M_{e\tau}=\pm M_{e\mu}$ and find allowed regions for $\sin\vartheta_{13}$ in each texture.

\section{\label{sec:3}Specific Textures}
Now, the determinant zero condition is used to ensure the presence of one massless neutrino, which allows the only two cases to be compatible with the observation: $(m_1, m_2, m_3)$=$(0,\pm\sqrt{\Delta m_\odot^2}, \pm\sqrt{\Delta m_{atm}^2 + \Delta m_\odot^2})$ leading to $\vert m_3\vert\gg\vert m_2\vert$ and $(m_1, m_2, m_3)$ = $(\pm\sqrt{\Delta m_{atm}^2},\pm\sqrt{\Delta m_{atm}^2+\Delta m_\odot^2},0)$ leading to $\vert m_1\vert\sim\vert m_2\vert$. The case with $(0,m_2,m_3)$ corresponds $NMH$ while the case with $(m_1,m_2,0)$ corresponds to $IMH$.  In $NMH$, we find that
\begin{eqnarray}
	\tan^2\vartheta_{12} = \frac{\lambda_1}{\lambda_2},
\label{Eq:sin12Simplified}
\end{eqnarray}
to realize $m_1=0$ in Eq.(\ref{Eq:TwoMasses1-2Simplified}). 

By requiring in Eq.(\ref{Eq:MassMatrixElements}) that one of the matrix elements vanishes or that the relations of $M_{\mu\mu} = M_{\tau\tau}$ and $M_{e\tau} = \pm M_{e\mu}$ are satisfied, we obtain the following constraints for the ratio of $m_3/m_2$ in $NMH$ with $m_1=0$:
\begin{eqnarray}
M_{ee} = 0 &\rightarrow&   \frac{m_3}{m_2} = -\frac{s_{12}^2c_{13}^2}{s_{13}^2},
\hspace{2.36cm}
M_{e\mu} = 0 \rightarrow \frac{m_3}{m_2} = -\frac{s_{12}c_{12}c_{23}-s_{12}^2s_{23}s_{13}}{s_{23}s_{13}},
\nonumber \\
M_{e\tau} =  0 &\rightarrow&   \frac{m_3}{m_2}=\frac{s_{12}c_{12}s_{23}+s_{12}^2c_{23}s_{13}}{c_{23}s_{13}},
\quad
M_{\mu\mu} = 0 \rightarrow \frac{m_3}{m_2}=-\frac{c_{12}^2c_{23}^2+s_{12}^2s_{23}^2s_{13}^2-2s_{12}c_{12}s_{23}c_{23}s_{13}}{s_{23}^2c_{13}^2},
\nonumber \\
M_{\mu\tau} = 0 &\rightarrow& \frac{m_3}{m_2}=\frac{\left( c_{12}^2-s_{12}^2s_{13}^2\right) s_{23}c_{23}+s_{12}c_{12}\left(c_{23}^2-s_{23}^2\right)s_{13}}{s_{23}c_{23}c_{13}^2},
\nonumber \\
M_{\tau\tau} = 0 &\rightarrow& \frac{m_3}{m_2}=-\frac{c_{12}^2s_{23}^2+ s_{12}^2c_{23}^2s_{13}^2+2s_{12}c_{12}s_{23}c_{23}s_{13}}{c_{23}^2c_{13}^2},
\nonumber \\
M_{\mu\mu} = M_{\tau\tau} &\rightarrow& \frac{m_3}{m_2}=\frac{(c_{12}^2-s_{12}^2s_{13}^2)(c_{23}^2-s_{23}^2)-4s_{12}c_{12}s_{23}c_{23}s_{13}}{(c_{23}^2-s_{23}^2)c_{13}^2},
\nonumber \\
M_{e\tau} = \eta M_{e\mu} &\rightarrow& \frac{m_3}{m_2}=
\frac{s^2_{12}\left( c_{23}-\eta s_{23}\right)s_{13}+s_{12}c_{12}\left( \eta c_{23}+s_{23}\right)}{(c_{23}-\eta s_{23})s_{13}},
\label{Eq:elementsWithNormalMassHierarchy}
\end{eqnarray}
where $\eta=\pm 1$.  We can also obtain the following constraints for the ratio of $m_2/m_1$ in $IMH$ with $m_3=0$:
\begin{eqnarray}
M_{ee}  = 0 &\rightarrow&\frac{m_2}{m_1} = -\frac{c_{12}^2}{s_{12}^2}, 
\hspace{2.82cm}
M_{e\mu} = 0 \rightarrow  \frac{m_2}{m_1} =\frac{ s_{12}c_{12}c_{23} +c_{12}^2s_{23}s_{13}}{s_{12}c_{12}c_{23}-s_{12}^2s_{23}s_{13}}, 
\nonumber \\
M_{e\tau} =  0 &\rightarrow& \frac{m_2}{m_1} = \frac{s_{12}c_{12}s_{23} - c_{12}^2c_{23}s_{13}}{s_{12}c_{12}s_{23}+s_{12}^2c_{23}s_{13}},
\quad
M_{\mu\mu} = 0 \rightarrow \frac{m_2}{m_1}=-\frac{s_{12}^2c_{23}^2+c_{12}^2s_{23}^2s_{13}^2+2s_{12}c_{12}s_{23}c_{23}s_{13}}{c_{12}^2c_{23}^2+s_{12}^2s_{23}^2s_{13}^2-2s_{12}c_{12}s_{23}c_{23}s_{13}},
\nonumber \\
M_{\mu\tau} =  0 &\rightarrow& \frac{m_2}{m_1} =- \frac{\left( s_{12}^2-c_{12}^2s_{13}^2\right) s_{23}c_{23}-s_{12}c_{12}\left( c_{23}^2-s_{23}^2\right)s_{13}}{\left( c_{12}^2-s_{12}^2s_{13}^2\right) s_{23}c_{23}+s_{12}c_{12}\left( c_{23}^2-s_{23}^2\right)s_{13}},
\nonumber \\
M_{\tau\tau}=  0 &\rightarrow&\frac{m_2}{m_1}=-\frac{s_{12}^2s_{23}^2+c_{12}^2c_{23}^2s_{13}^2-2s_{12}c_{12}s_{23}c_{23}s_{13}}{c_{12}^2s_{23}^2 + s_{12}^2c_{23}^2s_{13}^2+2s_{12}c_{12}s_{23}c_{23}s_{13}},
\nonumber \\
M_{\mu\mu} = M_{\tau\tau} &\rightarrow& \frac{m_2}{m_1}=-\frac{\left( s_{12}^2-c_{12}^2s_{13}^2\right)\left( c_{23}^2-s_{23}^2\right)+4s_{12}c_{12}s_{23}c_{23}s_{13}}{\left(c_{12}^2-s_{12}^2s_{13}^2\right)\left( c_{23}^2-s^2_{23}\right)-4s_{12}c_{12}s_{23}c_{23}s_{13}},
\nonumber \\
M_{e\tau} = \eta M_{e\mu} &\rightarrow& \frac{m_2}{m_1}=-
\frac{c^2_{12}\left( c_{23}-\eta s_{23}\right)s_{13}-s_{12}c_{12}\left( \eta c_{23}+s_{23}\right)}{s^2_{12}\left( c_{23}-\eta s_{23}\right)s_{13}+s_{12}c_{12}\left( \eta c_{23}+s_{23}\right)}.
\label{Eq:elementsWithInvertedMassHierarchy}
\end{eqnarray}
Using the experimental data of $\Delta m_{atm,\odot}^2$, we find the upper and lower limits of $m_3^2/m_2^2$ for $NMH$ and of $m_2^2/m_1^2$ for $IMH$ as shown in Table \ref{Tab:m3m2_m2m1}. These ratios are calculated from Eqs.(\ref{Eq:elementsWithNormalMassHierarchy}) and (\ref{Eq:elementsWithInvertedMassHierarchy}) with the observed mixing angles in Eq.(\ref{Eq:Data}). As already pointed out in the Introduction, the sign of $s_{23}$ is considered to obtain $s_{13}$ because $s_{13}$ is proportional to $Y=s_{23}M_{e\mu} +c_{23}M_{e\tau}$ as in Eq.(\ref{Eq:X-Y}),\footnote{We assume that $s^2_{23}<c^2_{23}$ and do not include the case with $s^2_{23}>c^2_{23}$ because the texture with interchange of $s_{23}\leftrightarrow c_{23}$ accompanied by the appropriate change of the sign of $s_{13}$ becomes identical to one of the existing textures.} where $Y$ can reduce its magnitude depending upon the sign of $s_{23}$.  From $\Delta m_{atm}^2$ and $\Delta m_\odot^2$, the allowed region of $m_3^2/m_2^2$ with $NMH$ and of $m_2^2/m_1^2$ with $IMH$ are computed to be $13.6 \le m_3^2/m_2^2 \le 89.9$ and $1.01 \leq  m_2^2/m_1^2 \leq 1.08$, respectively. 

We find the allowed textures with
\begin{itemize}
\item in the case of $NMH$, 
\begin{itemize}
\item $M_{ee}=0$, $M_{e\mu}=0$ and $M_{e\tau}=0$  as well as $M_{\mu\mu}=M_{\tau\tau}$ and $M_{e\tau}=-M_{e\mu}$ for $\sin\theta_{23}>0$
\item $M_{ee}=0$, $M_{e\mu}=0$ and $M_{e\tau}=0$ as well as $M_{\mu\mu}=M_{\tau\tau}$ and $M_{e\tau}=M_{e\mu}$ for $\sin\theta_{23}<0$
\end{itemize}
\item in the case of $IMH$, 
\begin{itemize}
\item 
$M_{e\mu}=0$ and $M_{\mu\mu}=0$ as well as $M_{\mu\mu}=M_{\tau\tau}$ and $M_{e\tau}=-M_{e\mu}$ for $\sin\theta_{23}>0$,
\item 
$M_{e\tau}=0$ and $M_{\tau\tau}=0$ as well as $M_{e\tau}=\pm M_{e\mu}$ for $\sin\theta_{23}<0$,
\end{itemize}
\end{itemize}
from Table \ref{Tab:m3m2_m2m1}.  Other textures with one vanishing matrix element are excluded by the observed constraints on the masses and the mixing angles.  In $NMH$, the allowed textures have the property that $m_3/m_2$ gets increased either as $c^2_{23}\rightarrow s^2_{23}$ ($\sin^2 2\theta_{23}\rightarrow 1$) for $M_{\mu\mu}=M_{\tau\tau}$ or as $s_{13}\rightarrow 0$ for other textures to meet the experimentally allowed value of $m^2_3/m^2_2$. It is thus expected that $\sin^2 2\theta_{23}\approx 1.0$ in the texture with $M_{\mu\mu}=M_{\tau\tau}$.  The texture with two zeros such as $M_{e\mu}$=$M_{e\tau}$=0 is excluded simply because $\tan^2 \vartheta_{23} = -1$ is satisfied. Similarly for other excluded textures. 

\begin{figure}[!htbp]
\begin{flushleft}
\includegraphics*[60mm,197mm][300mm,278mm]{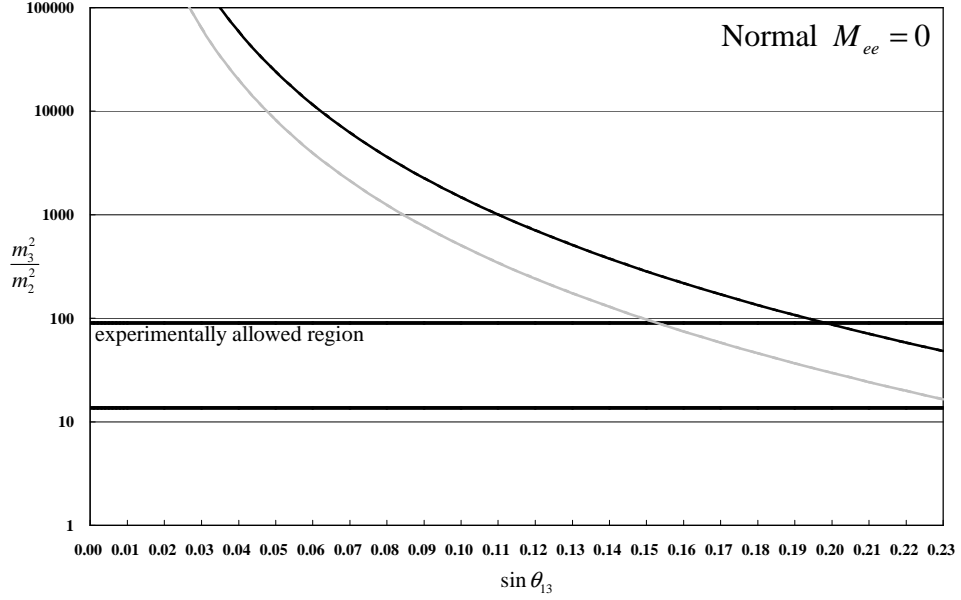}
\end{flushleft}
\caption{The prediction of $m^2_3/m^2_2$ for $NMH$: The area between the black curve (upper bound) and the gray curve (lower bound) is our prediction on $m^2_3/m^2_2$ in the texture with $M_{ee}=0$, which  does not depend on the sign of $\sin\vartheta_{23}$.  The area between two straight lines is the experimentally allowed region, which in turn determines the allowed values of $\sin\vartheta_{13}$.}
\label{Fig:ee}
\end{figure}

Shown in TABLE \ref{Tab:sin13} is the summary of our predictions on the values of $\sin\vartheta_{23}$, which are depicted in FIG.\ref{Fig:ee}-FIG.\ref{Fig:etau_-emu-inv}.  The effective neutrino mass $m_{\beta\beta}$ used in the detection of the absolute neutrino mass \cite{AbsoluteMass} is computed in each texture and the lower and upper bounds on $m_{\beta\beta}$ (=$M_{ee}$) in the present discussions are tabulated in TABLE \ref{Tab:ee}.  Also computed in TABLE \ref{Tab:NMH_M} and TABLE \ref{Tab:IMH_M} is each element of $M$ for the typical values of mixing angles such as $\sin^22\vartheta_{12} = 0.80$ and $\sin^22\vartheta_{12} = 0.98$ to see how $\sin^2\theta_{13}\ll 1$ as well as some hierarchies among $\lambda_{1,2,3}$ are realized.  The following characteristic features of the textures are found from these tables and figures:
\begin{enumerate}
\item The announced relations of $M_{\mu\mu}+M_{\tau\tau}\sim 2\sigma M_{\mu\tau}$ and $\vert M_{\mu\mu}+M_{\tau\tau}\vert \gg \vert M_{ee}\vert$ in $NMH$ and of $M_{\mu\mu}+M_{\tau\tau}\sim -2\sigma M_{\mu\tau}$ and $M_{\mu\mu}+M_{\tau\tau} \sim  \pm M_{ee}$ in $IMH$ with $m_1 \sim \pm m_2$ are satisfied. 
\item There are additional hierarchies of $\vert M_{e\mu}\vert, \vert M_{e\tau}\vert \ll \vert M_{\mu\mu}+M_{\tau\tau}\vert$ in $NMH$ and of $\vert M_{ee}\vert \gg \vert \sigma M_{e\mu}+M_{e\tau}\vert$ in $IMH$. Furthermore, $\vert M_{e\mu}-\sigma M_{e\tau}\vert \gg \vert \sigma M_{e\mu} +M_{e\tau}\vert$ expected in C2) is satisfied in the $IMH$-textures with $M_{\mu\mu}=0$ ($\sigma=1$), $M_{\tau\tau}=0$ ($\sigma=-1$), $M_{\mu\mu}=M_{\tau\tau}$ ($\sigma=1$) and $M_{e\tau}=M_{e\mu}$ ($\sigma=-1$), all  exhibiting $m_1\sim -m_2$.
\item The suppressed $\tan 2\vartheta_{13}$ is realized by the dominated magnitude of $\lambda_3$ ($\sim M_{\mu\mu}+M_{\tau\tau}+ 2\sigma M_{\mu\tau}$) for $NMH$ and of $M_{ee}$ for $IMH$ (because of $\lambda_3\sim 0$ in Eq.(\ref{Eq:MixingAngles13IMH})). The more suppressed values of $\sin\vartheta_{13}\mapleq 0.05$ are obtained due to $s_{23}M_{e\mu}+c_{23}M_{e\tau}\sim 0$, which is possible to occur in the textures with $M_{\mu\mu}=M_{\tau\tau}$ (only for $IMH$) and $M_{e\tau}=\pm M_{e\mu}$.
\item In $NMH$, $\sin^2 2\vartheta_{23} =0.998$ is selected as a typical value in the texture with $M_{\mu\mu}=M_{\tau\tau}$, which requires that $c^2_{23}\rightarrow s^2_{23}$ in the denominator of $m_2/m_1$ in Eq.(\ref{Eq:elementsWithInvertedMassHierarchy}) since, in this limit, $m_2/m_1$ approaches to 1 by passing the experimentally allowed $m_2/m_1$, 
\item In $IMH$, the small magnitude of $\sin\vartheta_{13}$ calls for the suppressed and almost vanishing $M_{e\mu}$ ($M_{e\tau}$) for the texture with $M_{e\tau}=0$ ($M_{e\mu}=0$).
\item In $IMH$, the relatively small magnitude of $M_{ee}$ can enhance $\sin\vartheta_{13}(\mapgeq 0.15)$ as in Eq.(\ref{Eq:MixingAngles13IMH}) in the textures with $M_{\mu\mu}=0$, $M_{\tau\tau}=0$ and $M_{e\mu}=M_{e\tau}$.  The texture with $M_{\mu\mu}=M_{\tau\tau}$ that also shows the relatively small magnitude of $M_{ee}$ receives the cancellation in $s_{23}M_{e\mu} +c_{23}M_{e\tau}(=0.002)$ to yield $\sin\vartheta_{13}=0.037$.
\item In $IMH$, the source of $m_{1,2}$ comes either from $\lambda_1\sim \lambda_2$ giving $m_2 \sim m_1$ for the textures with $M_{e\mu,e\tau}=0$ and $M_{e\tau}=\pm M_{e\mu}$ or from $X/\sin 2\vartheta_{12}$ with $\lambda_1+\lambda_2\sim 0$  giving $m_2 \sim - m_1$ for others.  Two sources coexist in the texture with $M_{\mu\mu}=M_{\tau\tau}$, where the lower bound on $m_2/m_1$ arises from the two sources as in FIG.\ref{Fig:mumu_tautau-inv}.
\item In $IMH$, the upper bounds on $m^2_2/m^2_1$ in FIG.\ref{Fig:mumu_tautau-inv} and FIG.\ref{Fig:etau_-emu-inv} for $\sin\vartheta_{23}>0$ and in FIG.\ref{Fig:etau_+emuu-inv} for $\sin\vartheta_{23}<0$ are very steep because, in Eq.(\ref{Eq:elementsWithInvertedMassHierarchy}), either $c^2_{23}-s^2_{23}$ in $M_{\mu\mu}=M_{\tau\tau}$ or $\eta c_{23}+s_{23}$ in $M_{e\tau}=\sigma M_{e\mu}$ can compete with $s_{13}$ so that the cancellation occurs in the denominators around $c^2_{23}\sim s^2_{23}$ ($M_{\mu\mu}=M_{\tau\tau}$) and $\eta c_{23}\sim -s_{23}$ ($M_{e\tau}=\sigma M_{e\mu}$) to give the rapid rise of $m^2_2/m^2_1$.
\item The effective neutrino mass of $m_{\beta\beta}$ is at most 0.008 eV for $NMH$ and  at most 0.07 eV for $IMH$.
\end{enumerate}
Some of these features are based on the specific values of mixing angles listed in the tables.  However, some of the constraints are generally satisfied by Eq.(\ref{Eq:MassMatrixElements}) that do not depend on the details of models and it has been checked that these features are shared by other allowed set of values of mixing angles.

Let us finally comment on how to realize $\sin^22\vartheta_{12}\gg\sin^2\vartheta_{13}$. In $NMH$, we have obtained the approximate relations of
\begin{eqnarray}
&&M_{\mu\mu}+M_{\tau\tau}\sim 2\sigma M_{\mu\tau},
\label{Eq:NMHApprox-1} \\
&&\vert M_{\mu\mu}+M_{\tau\tau}\vert\gg\vert M_{ee}\vert,
\label{Eq:NMHApprox-2} \\
&&\vert M_{\mu\mu}+M_{\tau\tau}\vert\gg\vert \sigma M_{e\mu}+M_{e\tau}\vert.
\label{Eq:NMHApprox-3}
\end{eqnarray}
Since Eq.(\ref{Eq:MixingAngles12-2}) leads to
\begin{eqnarray}
&& \tan2\vartheta_{12} \sim \frac{2\sqrt{2}\left(M_{e\mu}-\sigma M_{e\tau}\right)}{M_{\mu\mu} + M_{\tau\tau} - 2 \sigma M_{\mu\tau} },
\end{eqnarray}
from Eq.(\ref{Eq:NMHApprox-2}) and $\vert\tan 2\vartheta_{12}\vert \gg 1$ arises because of Eq.(\ref{Eq:NMHApprox-1}); therefore, the constraint of C1) works.  We also find from Eq.(\ref{Eq:NMHApprox-2}) that Eq.(\ref{Eq:MixingAngles13-0}) leads to
\begin{eqnarray}
\tan2\vartheta_{13} \sim \frac{\sqrt{2}\left(\sigma M_{e\mu} +M_{e\tau}\right)}{ M_{\mu\mu} + M_{\tau\tau}},
\label{Eq:MixingAngles13-000}
\end{eqnarray}
which gives $\vert \tan\vartheta_{13}\vert\ll 1$ because of Eq.(\ref{Eq:NMHApprox-3}).  In $IMH$, we have obtained the approximate relations of
\begin{eqnarray}
&&M_{\mu\mu}+M_{\tau\tau} \sim -2\sigma M_{\mu\tau},
\label{Eq:IMHApprox-1} \\
&&\vert \sigma M_{e\mu}+M_{e\tau}\vert\ll\vert M_{ee}\vert,
\label{Eq:IMHApprox-2}
\end{eqnarray}
as well as
\begin{eqnarray}
&&M_{\mu\mu}+M_{\tau\tau} \sim M_{ee},
\label{Eq:IMHApprox-3}
\end{eqnarray}
for $m_1\sim m_2$, and
\begin{eqnarray}
&&M_{\mu\mu}+M_{\tau\tau} \sim -M_{ee} \sim -2\sigma M_{\mu\tau},
\label{Eq:IMHApprox-4} \\
&&\vert M_{e\mu}-\sigma M_{e\tau}\vert \gg \vert \sigma M_{e\mu}+M_{e\tau}\vert,\label{Eq:IMHApprox-5}
\end{eqnarray}
for $m_1\sim -m_2$.  For $m_1\sim m_2$,
\begin{eqnarray}
&& \tan2\vartheta_{12} \sim \frac{\sqrt{2}\left(M_{e\mu}-\sigma M_{e\tau}\right)}{M_{\mu\mu} + M_{\tau\tau} - M_{ee}},
\end{eqnarray}
from Eq.(\ref{Eq:IMHApprox-1}), leading to $\vert\tan 2\vartheta_{12}\vert \gg 1$ because of Eq.(\ref{Eq:IMHApprox-3}).  The constraint of C1) works for $m_1\sim m_2$. From Eq.(\ref{Eq:MixingAngles13IMH}),
\begin{eqnarray}
&& \tan2\vartheta_{13} \sim -\frac{\sqrt{2}\left(\sigma M_{e\mu}+M_{e\tau}\right)}{M_{ee} },
\label{Eq:MixingAngles13IMH-1}
\end{eqnarray}
is obtained and $\vert\tan 2\vartheta_{13}\vert \ll 1$ arises because of Eq.(\ref{Eq:IMHApprox-2}) in both $m_1 \sim m_2$ and $m_1 \sim -m_2$.  On the other hand, for $m_1\sim -m_2$, we obtain that
\begin{eqnarray}
&& \tan2\vartheta_{12} \sim -\frac{\sqrt{2}\left(M_{e\mu}-\sigma M_{e\tau}\right)}{2 M_{ee} },
\end{eqnarray}
from Eq.(\ref{Eq:IMHApprox-4}). By comparing it with Eq.(\ref{Eq:MixingAngles13IMH-1}), we observe that $\vert \tan 2\vartheta_{12}\vert \gg\vert \tan 2\vartheta_{13}\vert$ is due to Eq.(\ref{Eq:IMHApprox-5}). The constraint of C2) works for $m_1\sim -m_2$. Therefore, the constraints of C1) and C2), repectively, work for $NMH$ and $IMH$ with $m_1\sim m_2$ and for $IMH$ with $m_1\sim -m_2$ to yield $\sin^22\vartheta_{12} \gg\sin^2\vartheta_{13}$, which also calls for the additional hierarchy of either $\vert M_{\mu\mu}+M_{\tau\tau}\vert\gg\vert s_{12} M_{e\mu}+c_{12}M_{e\tau}\vert$ ($NMH$), or $\vert M_{ee}\vert \gg \vert c_{23}M_{e\mu}+s_{23} M_{e\tau}\vert$ ($IMH$) to suppress the magnitude of $\sin^2\vartheta_{13}$ itself. Of course, the cancellation of $c_{23}M_{e\mu}+s_{23} M_{e\tau}\sim 0$ can be one of the sources of this hierarchy.

\section{\label{sec:4}Summary and Discussions}
We have suggested that the almost maximal atmospheric neutrino mixing occurs if
\begin{eqnarray}
	M_{e\tau}\approx -\sigma M_{e\mu}~{\rm and/or}~M_{\tau\tau} \approx M_{\mu\mu},
\label{Eq:Conditions1-Summary}
\end{eqnarray}
as $\sin\vartheta_{13}\rightarrow 0$ and if
\begin{eqnarray}
	M_{e\tau}\approx \sigma M_{e\mu}~{\rm and}~M_{\tau\tau} \approx M_{\mu\mu},
\label{Eq:Conditions2-Summary}
\end{eqnarray}
irrespective of the magnitude of $\sin\vartheta_{13}$.  The useful formulas are derived to calculate the mixing angles of $\theta_{12,13}$ in terms of the flavor neutrino masses, $M_{ij}$ for $i,j=e,\mu,\tau$:
\begin{eqnarray}
&& \tan2\vartheta_{13} = \frac{2\left(s_{23}M_{e\mu} +c_{23}M_{e\tau}\right)}{s_{23}^2 M_{\mu\mu} + c_{23}^2 M_{\tau\tau} + 2 s_{23} c_{23}M_{\mu\tau} - M_{ee}},
\label{Eq:MixingAngles2}
\end{eqnarray}
and 
\begin{eqnarray}
&& \tan2\vartheta_{12} \approx \frac{2\left(s_{23}M_{e\tau} -c_{23}M_{e\mu}\right)}{c_{23}^2 M_{\mu\mu} + s_{23}^2 M_{\tau\tau} - 2 s_{23} c_{23}M_{\mu\tau} - M_{ee}},
\label{Eq:MixingAngles3}
\end{eqnarray}
as far as $\sin^2\vartheta_{13}\approx 0$. We have suggested the mechanism to obtain the suppressed $\sin\vartheta_{13}$ based on
\begin{itemize}
\item the hierarchy of $\vert M_{ee,\mu\mu,\mu\tau{~{\rm and/or}~}\tau\tau}\vert \gg \vert\sigma M_{e\mu}+M_{e\tau}\vert $,
\item the cancellation of $s_{23}M_{e\mu} +c_{23}M_{e\tau}\sim 0$ (for more suppressed $\sin\vartheta_{13}$),
\end{itemize}
and also to yield $\sin^22\vartheta_{12} \gg\sin^2\vartheta_{13}$ based on
\begin{itemize}
\item the approximate equality of $c_{23}^2 M_{\mu\mu} + s_{23}^2 M_{\tau\tau} \approx 2 s_{23} c_{23}M_{\mu\tau} + M_{ee}$,
\item the hierarchy of $\vert c_{23}M_{e\mu}-s_{23}M_{e\tau}\vert \gg \vert s_{23}M_{e\mu} +c_{23}M_{e\tau}\vert$.
\end{itemize}

Considering various general constraints on flavor neutrino masses shown in TABLE \ref{Tab:Masses}, we reach the plausible textures at the ``zero"-th order: 
\begin{eqnarray}
M^{NMH} =
    \left( 
    \begin{array}{ccc}
    \epsilon_1 & \epsilon_2 & \epsilon_3\\
   \epsilon_2 & d & \sigma d\\
    \epsilon_3 & \sigma d & d\\
    \end{array} 
    \right), 
    \quad
M^{IMH} =
    \left( 
    \begin{array}{ccc}
    \pm 2 d & b & -\sigma b\\
   b & d & -\sigma d\\
    -\sigma b & -\sigma d & d\\
    \end{array} 
    \right)~({\rm for}~m_1\sim\pm m_2),
\label{Eq:GrossStrctureSummery}
\end{eqnarray}
where $\vert\epsilon_{1,2,3}\vert\ll \vert d\vert$ in $M^{NMH}$ and $\sigma = \pm 1$ for $\sin\vartheta_{23} = \sigma/\sqrt{2}$. It should be noted that the pattern in $M^{NMH}$ may reflect a symmetry based on the approximate conservation of the electron number, $L_e$. Namely, $\epsilon_1$ has $L_e=2$, $\epsilon_{2,3}$ have $L_e=1$ and $d$ has $L_e=0$ \cite{eNumber,mu-tau2}. In the case that the conservation of $L_e$ is perturbatively violated by an interaction of $L_e = \pm 1$ characterized by an appropriate parameter of $\xi$ with $\vert \xi \vert \ll 1$ \cite{mu-tau2}, it is not absurd to expect $ \epsilon_1 \propto \xi^2$, $ \epsilon_{2,3} \propto \xi$  and $ d \propto \xi^0$, which explain the required order in $M^{NMH}$.  

In specific models with one massless neutrino, the above-mentioned mechanisms have been confirmed to yield the consistent magnitudes of $\sin\vartheta_{13}$ and $\sin^22\vartheta_{12}$ in various textures.  The allowed textures are found to exhibit the following patterns of the flavor neutrino masses:
\begin{itemize}
\item For $NMH$, either $M_{ee}$, $M_{e\mu}$ or $M_{e\tau}$ can vanish and 
\begin{eqnarray}
\vert M_{\mu\mu}+M_{\tau\tau}\vert\gg \vert M_{ee}\vert, \vert \sigma M_{e\mu}+M_{e\tau}\vert, \quad
M_{\mu\mu}+M_{\tau\tau}\sim 2\sigma M_{\mu\tau},
\label{Eq:SummaryNMH}
\end{eqnarray}
are satisfied, and
\item For $IMH$,
either $M_{e\mu}$, $M_{e\tau}$, $M_{\mu\mu}$ and $M_{\tau\tau}$ can vanish and 
\begin{eqnarray}
\vert M_{ee}\vert \sim  \vert M_{\mu\mu}+M_{\tau\tau}\vert \gg \vert \sigma M_{e\mu}+M_{e\tau}\vert, \quad
M_{\mu\mu}+M_{\tau\tau}\sim -2\sigma M_{\mu\tau},
\label{Eq:SummaryIMH}
\end{eqnarray}
are satisfied,
\end{itemize}
where $\sigma$ reflects the sign of $\sin\vartheta_{23}$. As a result, the constraint of C1) works in the textures for $NMH$ and $IMH$ with $m_1\sim m_2$ while the constraint of C2) works in the textures for $IMH$ with $m_1\sim -m_2$.  It can be generally expected that these relations are satisfied in any other models.

The effective neutrino mass of $m_{\beta\beta}$ is found to be bounded as $\vert m_{\beta\beta} \vert\mapleq 0.008$ eV for $NMH$ and $\vert m_{\beta\beta} \vert\mapleq 0.07$ eV for $IMH$. This predicted maximal value of $m_{\beta\beta}$ is outside of 0.22 eV$\mapleq \vert m_{\beta\beta} \vert\mapleq$1.6 eV \cite{AbsoluteMass} obtained from the Heidelberg-Moscow experiment \cite{ee-exp}.  Therefore, none of the textures examined in this article is compatible with this observation if it is confirmed by other experiments.  Which are appropriate textures if future experiments \cite{theta-13-ex} point to, say, $\sin\theta_{13}$ =0.01, 0.1 or 0.2? The results are shown in TABLE \ref{Tab:Summary}, which can be readily read off from the figures.  At first glance, the textures suggested by the $\mu$-$\tau$ symmetry have any values of $\sin\vartheta_{13}$.  It is because the $\mu$-$\tau$ symmetry may represent an underlying symmetry of neutrino oscillations if it exists at all.  On the other hand, the textures with one vanishing matrix element need to be strongly constrained to explain the observed data.  The larger values of $\sin\vartheta_{13}$ are favored unless the appropriate nonvanishing elements are extremely suppressed.  

\appendix
\section{\label{sec:Appendix}Neutrino Masses and Mixing Angles}
After a simple diagonalization of $M$ by $U$, one can find the following conditions:
\begin{eqnarray}
&& c_{12} \Delta _1  - s_{12}c_{13} \left( {s_{13}  X  +  \Delta _2 } \right) = 0,
\quad
s_{12} \Delta _1  + c_{12} c_{13}\left( {s_{13}  X  + \Delta _2 } \right) = 0,
\noindent \\
&&s_{12} \left[ {c_{12} \lambda _1  - s_{12} \left( {c^2_{13}X  - s_{13} \Delta _2 } \right)} \right] + c_{12} \left[ {c_{12} \left( {c^2_{13}X  - s_{13} \Delta _2 } \right) - s_{12} \lambda _2 } \right] = 0
\end{eqnarray}
with
\begin{eqnarray}
&&
\Delta _1  = c_{13} s_{13} \left( {M_{ee} - \lambda _3 } \right) + \left( {c_{13}^2  - s_{13}^2 } \right)\left( {s_{23} M_{e\mu} + c_{23} M_{e\tau}} \right),
\nonumber \\
&&
\Delta _2  = \left( {c_{23}^2  - s_{23}^2 } \right)M_{\mu\tau} + s_{23} c_{23} \left( {M_{\mu\mu} - M_{\tau\tau}} \right),
\label{Eq:Delta}
\end{eqnarray}
where
\begin{eqnarray}
&&
\lambda_1 = c_{13}^2 M_{ee} + s_{13}^2 \lambda_3 - 2s_{13}c_{13}Y,
\quad
\lambda_2 = c_{23}^2 M_{\mu\mu} + s_{23}^2 M_{\tau\tau} - 2 s_{23} c_{23}M_{\mu\tau}, 
\nonumber \\
&&
\lambda_3 = s_{23}^2 M_{\mu\mu} + c_{23}^2 M_{\tau\tau} + 2 s_{23} c_{23}M_{\mu\tau},
\label{Eq:lambdas}\\
&&
X = \frac{ c_{23}M_{e\mu} -s_{23}M_{e\tau} }{c_{13}},
\quad
Y = s_{23}M_{e\mu} +c_{23}M_{e\tau}.
\label{Eq:X-Y}
\end{eqnarray}
The mixing angles of $\vartheta_{12}$ and $\vartheta_{13}$ are then expressed as:
\begin{eqnarray}
&&
\tan2\vartheta_{12} = \frac{2X}{\lambda_2 - \lambda_1}, 
\quad
\tan2\vartheta_{13} = \frac{2Y}{ \lambda_3 - M_{ee} },
\label{Eq:MixingAngles}
\end{eqnarray}
together with
\begin{eqnarray}
&&
M_{\tau\tau} = M_{\mu\mu} + 2\frac{\cos 2\vartheta _{23}M_{\mu\tau} + s_{13}X}{\sin 2\vartheta _{23} }.
\label{Eq:f-d}
\end{eqnarray}
The neutrino masses of $m_{1,2,3}$ are given by\footnote{In Ref.\cite{mu-tau2}, Eqs.(4) and (6) should read the corresponding expressions in Eqs.(\ref{Eq:lambdas})-(\ref{Eq:ThreeMasses}) of this article.}
\begin{eqnarray}
	m_1 &=& c_{12}^2 \lambda_1 + s_{12}^2 \lambda_2 - 2s_{12}c_{12}X, \quad
	m_2 = s_{12}^2 \lambda_1 + c_{12}^2 \lambda_2 + 2s_{12}c_{12}X,
\nonumber \\
	m_3 &=& s_{13}^2 M_{ee} + c_{13}^2 \lambda_3 + 2s_{13}c_{13}Y,
\label{Eq:ThreeMasses}
\end{eqnarray}
and $m_{1,2}$ can be converted into
\begin{eqnarray}
	m_1 &=& \frac{\lambda_1 + \lambda_2}{2}-\frac{X}{\sin2\vartheta_{12}}, \quad
	m_2 = \frac{\lambda_1 + \lambda_2}{2}+\frac{X}{\sin2\vartheta_{12}},
\label{Eq:TwoMasses1-2}
\end{eqnarray}
by using Eq.(\ref{Eq:MixingAngles}).  These expressions are further simplified into
\begin{eqnarray}
	m_1 &=& \frac{c^2_{12}\lambda_1-s^2_{12}\lambda_2}{c^2_{12}-s^2_{12}}, \quad
	m_2  = \frac{c^2_{12}\lambda_2-s^2_{12}\lambda_1}{c^2_{12}-s^2_{12}}.
\label{Eq:TwoMasses1-2Simplified}
\end{eqnarray}
These relations of masses and mixings enable us to discuss constraints on flavor neutrino masses.  


\begin{table}[!htbp]
    \caption{\label{Tab:Masses}General constraints on flavor neutrino masses of $M_{ij}$ for $i,j=e,\mu,\tau$ that provide $\sin^22\vartheta_{atm}\sim 1$, $\sin^2\vartheta_{13}\ll 1$, $\sin^22\vartheta_\odot\gg\sin^2\vartheta_{13}$, and $\Delta m^2_{atm}\gg\Delta m^2_\odot$, where $\cos\vartheta_{23}\sim \sigma\sin\vartheta_{23}\sim 1/\sqrt{2}$ ($\sigma=\pm 1$). In $\Delta m^2_{atm}\gg\Delta m^2_\odot$, $\vert M_{\mu\mu}+M_{\tau\tau}\vert\gg\vert M_{ee}\vert$ is satisfied if $M_{\mu\mu}+M_{\tau\tau}\sim2\sigma M_{\mu\tau}+2M_{ee}$ is chosen for $\sin^22\vartheta_\odot\gg\sin^2\vartheta_{13}$.}
    \begin{center}
    \begin{tabular}{|c|c||c|c|c|c|c|}
    \hline
\multicolumn{2}{|c||}{assumption}&
\multicolumn{2}{c|}{$\sin^22\vartheta_{atm}\sim 1$}&
$\sin^2\vartheta_{13}\ll 1$&
$\sin^22\vartheta_\odot\gg\sin^2\vartheta_{13}$&
$\Delta m^2_{atm}\gg\Delta m^2_\odot$
\\ \hline

\multicolumn{2}{|c||}{}&
\multicolumn{3}{c|}{}&
$M_{\mu\mu}+M_{\tau\tau}\sim 2\sigma M_{\mu\tau}+2M_{ee}$&
$M_{\mu\mu}+M_{\tau\tau}\sim 2\sigma M_{\mu\tau}$

\\

\multicolumn{2}{|c||}{$NMH$}&
\multicolumn{3}{c|}{$M_{e\tau}\sim -\sigma M_{e\mu}$}&
and/or&
($\ast$)$\vert M_{\mu\mu}+M_{\tau\tau}\vert\gg\vert M_{ee}\vert$

\\

\multicolumn{2}{|c||}{}&
\multicolumn{3}{c|}{and/or}&
$\vert M_{e\mu}-\sigma M_{e\tau}\vert\gg\vert \sigma M_{e\mu}+M_{\mu\tau}\vert$&

\\ \cline{1-3}\cline{6-7}
&
$m_1\sim m_2$&
$M_{e\tau}\sim \sigma M_{e\mu}$&
$M_{\mu\mu}\sim M_{\tau\tau}$&
$\vert M_{ee,\mu\mu,\mu\tau~{\rm and/or}~\tau\tau}\vert$&
$M_{\mu\mu}+M_{\tau\tau}\sim 2\sigma M_{\mu\tau}+2M_{ee}$&
$M_{\mu\mu}+M_{\tau\tau}\sim -2\sigma M_{\mu\tau}$

\\ \cline{2-2}\cline{6-6}

$IMH$&
$m_1\sim -m_2$&
and&
&
$\gg\vert \sigma M_{e\mu}+M_{\mu\tau}\vert$&
$\vert M_{e\mu}-\sigma M_{e\tau}\vert\gg\vert \sigma M_{e\mu}+M_{\mu\tau}\vert$&$M_{\mu\mu}+M_{\tau\tau}\sim 2\sigma M_{\mu\tau}\pm 2M_{ee}$

\\
&
&
$M_{\mu\mu}\sim M_{\tau\tau}$&
&
&
&
($\ast$)$\pm$ for $m_1 \sim\pm m_2$

\\ \hline
    \end{tabular}
    \end{center}
\end{table}

\begin{table}[!htbp]
    \caption{\label{Tab:m3m2_m2m1}The upper and lower limits of $m_3^2/m_2^2$ with the normal mass hierarchy ($NMH$) and of $m_2^2/m_1^2$ with inverted mass hierarch ($IMH$) in models based on $\det(M)=0$ with specific textures, where the mass ratios are calculated from Eqs.(\ref{Eq:elementsWithNormalMassHierarchy}) and (\ref{Eq:elementsWithInvertedMassHierarchy}) with the observed mixing angles in Eq.(\ref{Eq:Data}).  The values in the parentheses are those for $\sin\theta_{23} < 0$.}
    \begin{center}
    \begin{tabular}{|c|c|c|c|c|}
    \hline
     assumption &\multicolumn{2}{c|}{$m_3^2/m_2^2$ ($NMI$)}  &\multicolumn{2}{c|}{$m_2^2/m_1^2$ ($IMH$)}\\ \hline 
     bound   &lower          & upper         & lower          & upper\\ 
    \hline
    $M_{ee}=0$       &$16.5(16.5)$ & $\infty(\infty)$ & $2.49(2.49)$ & $11.6(11.6)$ \\
    $M_{e\mu}=0$     &$2.55(4.21)$  & $\infty(\infty)$ & $1.00(0.26)$ & $2.64(1.00)$ \\
    $M_{e\tau}=0$    &$2.53(1.29)$  & $\infty(\infty)$ & $0.14(1.00)$ & $1.00(3.54)$ \\
    $M_{\mu\mu}=0$   &$0.19(0.82)$  & $1.91(3.03)$   & $0.09(0.09)$ & $2.48(0.40)$ \\
    $M_{\mu\tau}=0$  &$0.38(0.38)$  & $0.74(0.65)$   & $0.09(0.06)$ & $0.40(0.64)$ \\
    $M_{\tau\tau}=0$ &$0.12(0.04)$  & $1.06(0.59)$   & $0.09(0.09)$ & $0.04(4.56)$ \\
\hline
    $M_{\mu\mu}=M_{\tau\tau}$ 
                   &$\sim 0(0.36)$  & $1.39\times 10^4(1.42\times 10^4)$   & $0.09(\sim 0)$ & $\sim \infty(0.37)$ \\
    $M_{e\tau}=M_{e\mu}$ 
                   &$166(0.42)$  & $\infty(\infty)$   & $0.85(\sim 0)$ & $1.00(11.4)$ \\
    $M_{e\tau}=-M_{e\mu}$ 
                   &$\sim 0(154)$  & $\infty(\infty)$   & $1.00(1.00)$ & $\sim \infty (1.17)$ \\
    \hline
    \end{tabular}
    \end{center}
\end{table}

\begin{table}[!htbp]
    \caption{\label{Tab:sin13}The predictions of $\sin\vartheta_{13}$ depending on the sign of $\sin\vartheta_{23}$ given by $\pm$ for $NMH$ and $IMH$, where the excluded cases are denoted by the dots. The absence of the allowed $M_{ee}$ for $IMH$ is due to Eq.(\ref{Eq:GeneralIMH})}
    \begin{center}
    \begin{tabular}{|c|c|c|c|c|}
    \hline
     assumption &\multicolumn{2}{c|}{$NMI$}  &\multicolumn{2}{c|}{$IMH$}\\ \hline 
     $\sin\vartheta_{23}$   &+          & $-$         & $+$          & -\\ 
    \hline
    $M_{ee}=0$       &\multicolumn{2}{c|}{$0.153\sim$} & $\cdots$ & $\cdots$ \\
    $M_{e\mu}=0$     &$0.043\sim 0.159$  & $0.045\sim 0.197$ & $0.001\sim 0.023$ & $\cdots$ \\
    $M_{e\tau}=0$    &$0.034\sim 0.147$  & $0.032\sim 0.119$ & $\cdots$ & $0.001\sim 0.017$ \\
    $M_{\mu\mu}=0$   &$\cdots$  & $\cdots$   & $0.115\sim$ & $\cdots$ \\
    $M_{\mu\tau}=0$  &$\cdots$  & $\cdots$   & $\cdots$  & $\cdots$ \\
    $M_{\tau\tau}=0$ &$\cdots$  & $\cdots$   & $\cdots$ & $0.086\sim$ \\
\hline
    $M_{\mu\mu}=M_{\tau\tau}$ 
                   &$0.006\sim$  & $0.009\sim$   & $0.0002\sim$ & $\cdots$ \\
    $M_{e\tau}=M_{e\mu}$ 
                   &$\cdots$  & $0.008\sim 0.021$   & $\cdots$ & $0.001\sim$ \\
    $M_{e\tau}=-M_{e\mu}$ 
                   &$0.007\sim 0.017$  & $\cdots$   & $\sim 0.003$ & $0.009\sim$ \\
    \hline
    \end{tabular}
    \end{center}
\end{table}

\begin{table}[!htbp]
    \caption{\label{Tab:ee}The same as in TABLE \ref{Tab:sin13} but for $\vert m_{\beta\beta}\vert $ in the unit of eV.}
    \begin{center}
    \begin{tabular}{|c|c|c|c|c|}
    \hline
     assumption &\multicolumn{2}{c|}{$NMI$}  &\multicolumn{2}{c|}{$IMH$}\\ \hline 
     $\sin\vartheta_{23}$   &+          & $-$         & $+$          & -\\ 
    \hline
    $M_{ee}=0$       &0 & 0 & 0 & 0 \\
    $M_{e\mu}=0$     &$0.001\sim 0.004$  & $0.002\sim 0.005$ & $0.03\sim 0.07$ & $\cdots$ \\
    $M_{e\tau}=0$    &$0.002\sim 0.004$  & $0.001\sim 0.004$ & $\cdots$ & $0.03\sim 0.07$ \\
    $M_{\mu\mu}=0$   &$\cdots$  & $\cdots$   & $0.008\sim 0.03$ & $\cdots$ \\
    $M_{\mu\tau}=0$  &$\cdots$  & $\cdots$   & $\cdots$  & $\cdots$ \\
    $M_{\tau\tau}=0$ &$\cdots$  & $\cdots$   & $\cdots$ & $0.007\sim 0.04$ \\
\hline
    $M_{\mu\mu}=M_{\tau\tau}$ 
                   &$0.00001\sim 0.004$  & $0.002\sim 0.008$   & $0.007\sim 0.07$ & $\cdots$ \\
    $M_{e\tau}=M_{e\mu}$ 
                   &$\cdots$  & $0.002\sim 0.004$   & $\cdots$ & $0.007\sim 0.04$ \\
    $M_{e\tau}=-M_{e\mu}$ 
                   &$0.002\sim 0.004$  & $\cdots$   & $0.03\sim 0.07$ & $0.03\sim 0.07$ \\
    \hline
    \end{tabular}
    \end{center}
\end{table}

\begin{table}[!htbp]
    \caption{\label{Tab:NMH_M}The list of calculated values of the flavor neutrino masses in the unit of $m_2$ in the allowed textures for $NMH$ with the specific mixing angles of $\sin 2\vartheta_{12}=0.80$, $\sin 2\vartheta_{23}=0.98$ (or $\sin 2\vartheta_{23}=0.998$) and the typical values of $\sin\vartheta_{13}$, where $\sigma$ (=$\pm 1$) stands for the sign of $\sin\vartheta_{23}$.}
    \begin{center}
    \begin{tabular}{|c|c|c|c||c|c|c|c|c|c|}
\hline
\multicolumn{1}{|c||}{condition}
&\multicolumn{1}{c|}{$\sin^2\vartheta_{12}$}
&\multicolumn{1}{c|}{$\sin^2\vartheta_{23}$}
&\multicolumn{1}{c||}{$\sin\vartheta_{13}$}  
&\multicolumn{1}{c|}{$\sigma$}
&\multicolumn{1}{c|}{$M_{ee}$} 
&\multicolumn{1}{c|}{$M_{e\mu}$} 
&\multicolumn{1}{c|}{$M_{e\tau}$} 
&\multicolumn{1}{c|}{$M_{\mu\mu}+M_{\tau\tau}$}  
&\multicolumn{1}{c|}{$2\sigma M_{\mu\tau}$} 
\\ \hline
$M_{ee}=0$ &0.80 & 0.98 &0.20  &+& 0 & $-$0.56 & $-$1.31 & $-$5.63 & $-$7.04
\\ \cline{5-10}
&&&&$-$&0 & 1.21 & $-$0.74  & $-$5.63 & $-$6.98
\\ \hline
$M_{e\mu}=0$ &0.80 & 0.98 &0.10&+ & 0.225 & 0 & $-$0.68  & $-$4.11 & $-$5.51
\\ \cline{5-10}
&&&&$-$&0.328 & 0 & 0.68 & 6.11 & 4.62 
\\ \hline
$M_{e\tau}=0$ &0.80 & 0.98 &0.09&+ & 0.31 & 0.59 & 0  & 5.28 & 3.78
\\ \cline{5-10}
&&&&$-$& 0.24 & 0.59 & 0  & $-$3.28 & $-$4.66
\\ \hline
$M_{\mu\mu}=M_{\tau\tau}$ &0.80 & 0.998 &0.22 &+& 0.076 & $-$0.30 & $-$0.95  & $-$2.95 & $-$4.40
\\ \cline{5-10}
&&&&$-$&0.52 & $-$0.44 & $-$1.09 & 5.84 & 4.40
\\ \hline
$M_{e\tau}=M_{e\mu}$ &0.80 & 0.98 &0.05 &+& $\cdots$ & $\cdots$ & $\cdots$  & $\cdots$ & $\cdots$
\\ \cline{5-10}
&&&&$-$& 0.28 & 0.32 & 0.32 & 7.36 & 5.85
\\ \hline
$M_{e\tau}=-M_{e\mu}$ &0.80 & 0.98 &0.05 &+& 0.28 & 0.32 & $-$0.32  & $-$5.36 & $-$6.74
\\ \cline{5-10}
&&&&$-$& $\cdots$ & $\cdots$ & $\cdots$ & $\cdots$ & $\cdots$
\\ \hline
    \end{tabular}
    \end{center}
\end{table}

\begin{table}[!htbp]
    \caption{\label{Tab:IMH_M}The same as in TABLE.\ref{Tab:NMH_M} but in the unit of $m_1$ for the allowed textures for $IMH$, where $\lambda_{1,2}$ are included to see $m_1\sim \pm m_2$ for $\lambda_1 \sim \pm\lambda_2$ and $\sin^2 2\theta_{12}=0.90$ is chosen in the texture with $M_{\mu\mu}=0$ instead of $\sin^2 2\theta_{12}=0.80$.}
    \begin{center}
    \begin{tabular}{|c|c|c|c||c|c|c|c|c|c|c|c|}
\hline
\multicolumn{1}{|c|}{condition} 
&\multicolumn{1}{c|}{$\sin^2\vartheta_{12}$}
&\multicolumn{1}{c|}{$\sin^2\vartheta_{23}$}
&\multicolumn{1}{c||}{$\sin\vartheta_{13}$}  
&\multicolumn{1}{c|}{$\sigma$}  
&\multicolumn{1}{c|}{$\lambda_1$} 
&\multicolumn{1}{c|}{$\lambda_2$} 
&\multicolumn{1}{c|}{$M_{ee}$} 
&\multicolumn{1}{c|}{$M_{e\mu}$} 
&\multicolumn{1}{c|}{$M_{e\tau}$} 
&\multicolumn{1}{c|}{$M_{\mu\mu}+M_{\tau\tau}$}  
&\multicolumn{1}{c|}{$2\sigma M_{\mu\tau}$} 

\\ \hline
      $M_{e\mu}=0$ &0.80 & 0.98 &0.100 &+ &1.01 &1.01 &1.01 & 0 & $-$0.01 & 1.01 & $-$1.00
\\ \hline
      $M_{e\tau}=0$  &0.80 & 0.98 &0.100&$-$&1.01 &1.02 &1.01 & 0.02 & 0 & 1.02 & $-$1.01
\\ \hline
      $M_{\mu\mu}=0$ &0.90 & 0.98 &0.190&+   &0.31 &$-$0.32 &0.30 & $-$0.75 & 0.57  & $-$0.31 & 0.38 
\\ \hline
      $M_{\tau\tau}=0$ &0.80 & 0.98 &0.210&$-$  &0.44 &$-$0.46 &0.42 & $-$0.61 & $-$0.65 & $-$0.45 & 0.43
\\ \hline
      $M_{\mu\mu}=M_{\tau\tau}$ &0.80 & 0.98&0.037 &+    &0.44 &$-$0.47 &0.44 & $-$0.70 & 0.58  & $-$0.47 & 0.48
\\ \hline
      $M_{e\tau}=M_{e\mu}$ &0.80 & 0.98 &0.145 &$-$&0.44 &$-$0.46 &0.43 & $-$0.63 & $-$0.63  & $-$0.45 & 0.43
\\ \hline
      $M_{e\tau}=-M_{e\mu}$ &0.80 & 0.98&0.001&+  &1.01 &1.02   &1.01 & 0.01 & $-$0.01   & 1.02 & $-$1.01
\\ \cline{2-12}
      &0.80 & 0.98&0.100&$-$  &1.00 &1.01 &0.99 & 0.07 & $-$0.07  & 1.02 & $-$0.99
\\ \hline
    \end{tabular}
    \end{center}
\end{table}

\begin{table}[!htbp]
    \caption{\label{Tab:Summary}The allowed textures denoted by $\bigcirc$ for $NMH$ and $IMH$ that give $\sin\vartheta_{13}=0.01,0.1,0.2$, where $\pm$ stand for $\sin\vartheta_{23}=\pm 1$ and the values of $A\equiv \sin^22\vartheta_{atm}$ and $S\equiv \sin^22\vartheta_\odot$ are shown if these angles are more constrained than those listed in Eq.(\ref{Eq:Data}).}
    \begin{center}
    \begin{tabular}{|c||c|c|c|c||c|c|c|c||c|c|c|c|}
    \hline
$\sin\vartheta_{13}$ &
\multicolumn{4}{c||}{$0.01$}  &
\multicolumn{4}{c||}{$0.1$}  &
\multicolumn{4}{c|}{$0.2$}\\
\hline 
assumption &
\multicolumn{2}{c|}{$NMI$}  &
\multicolumn{2}{c||}{$IMH$}  &
\multicolumn{2}{c|}{$NMI$}  &
\multicolumn{2}{c||}{$IMH$}  &
\multicolumn{2}{c|}{$NMI$}  &
\multicolumn{2}{c|}{$IMH$}\\ 
\hline 
type &
\multicolumn{1}{c|}{$+$}  &
\multicolumn{1}{c|}{$-$}  &
\multicolumn{1}{c|}{$+$}  &
\multicolumn{1}{c||}{$-$}  &
\multicolumn{1}{c|}{$+$}  &
\multicolumn{1}{c|}{$-$}  &
\multicolumn{1}{c|}{$+$}  &
\multicolumn{1}{c||}{$-$}  &
\multicolumn{1}{c|}{$+$}  &
\multicolumn{1}{c|}{$-$}  &
\multicolumn{1}{c|}{$+$}  &
\multicolumn{1}{c|}{$-$}\\
\hline
$M_{ee}=0$&
$\times$&$\times$&$\times$&$\times$&
$\times$&$\times$&$\times$&$\times$&
$\bigcirc$&$\bigcirc$&$\times$&$\times$
\\
$M_{e\mu}=0$&
$\times$&$\times$&$\bigcirc$&$\times$&
$\bigcirc$&$\bigcirc$&$\times$&$\times$&
$\times$&$\times$&$\times$&$\times$
\\
$M_{e\tau}=0$&
$\times$&$\times$&$\times$&$\bigcirc$&
$\bigcirc$&$\bigcirc$&$\times$&$\times$&
$\times$&$\times$&$\times$&$\times$
\\
&
&&&&
&$0.97\leq A$&&&
&&&
\\
$M_{\mu\mu}=0$&
$\times$&$\times$&$\times$&$\times$&
$\times$&$\times$&$\times$&$\times$&
$\times$&$\times$&$\bigcirc$&$\times$
\\
&
&&&&
&&&&
&&$0.86\leq S$&
\\
&
&&&&
&&&&
&&\qquad$\leq 0.92$&
\\
$M_{\mu\tau}=0$&
$\times$&$\times$&$\times$&$\times$&
$\times$&$\times$&$\times$&$\times$&
$\times$&$\times$&$\times$&$\times$
\\
$M_{\tau\tau}=0$&
$\times$&$\times$&$\times$&$\times$&
$\times$&$\times$&$\times$&$\bigcirc$&
$\times$&$\times$&$\times$&$\bigcirc$
\\
&
&&&&
&&&$A\leq 0.98$&
&&&$0.93\leq A$
\\
&
&&&&
&&&&
&&&$0.75\leq S$
\\
&
&&&&
&&&&
&&&\qquad $\leq 0.86$
\\
\hline
$M_{\mu\mu}=M_{\tau\tau}$& 
$\bigcirc$&$\bigcirc$&$\bigcirc$&$\times$&
$\bigcirc$&$\bigcirc$&$\bigcirc$&$\times$&
$\bigcirc$&$\bigcirc$&$\bigcirc$&$\times$
\\
& 
$A\sim 1$&$A\sim 1$&&&
$A\sim 1$&$A\sim 1$&&&
$A\sim 1$&$A\sim 1$&&
\\
$M_{e\tau}=M_{e\mu}$& 
$\times$&$\bigcirc$&$\times$&$\bigcirc$&
$\times$&$\times$&$\times$&$\bigcirc$&
$\times$&$\times$&$\times$&$\bigcirc$
\\
&
&$A\leq 0.98$&&$0.72\leq S$&
&&&$0.97\leq A$&
&&&$0.94\leq A$
\\
&
&&&\qquad$\leq 0.92$&
&&&&
&&&\qquad$\leq 0.99$
\\
$M_{e\tau}=-M_{e\mu}$& 
$\bigcirc$&$\times$&$\times$&$\times$&
$\times$&$\times$&$\times$&$\bigcirc$&
$\times$&$\times$&$\times$&$\bigcirc$
\\
&
$A\leq 0.97$&&&&
&&&&
&&&$0.97\leq A$
\\
\hline
    \end{tabular}
    \end{center}
\end{table}

\newpage
\noindent

\begin{figure}[!htbp]
\begin{flushleft}
\includegraphics*[60mm,197mm][300mm,278mm]{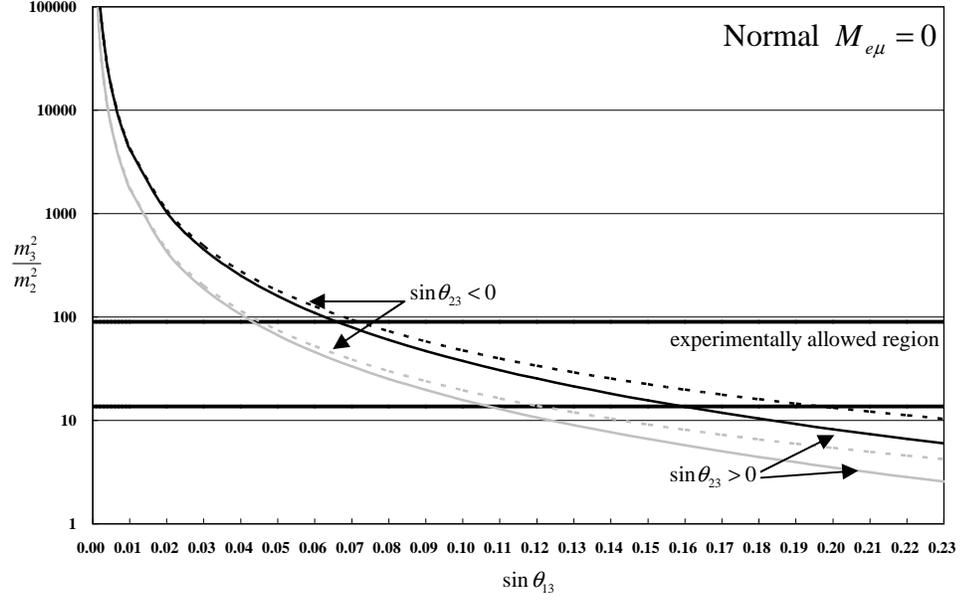}
\end{flushleft}
\caption{The same as in FIG.\ref{Fig:ee} but for $M_{e\mu}=0$. The solid (dashed) curves correspond to $\sin\theta_{23}>0$ ($\sin\theta_{23}<0$). The areas between the black curves (upper bounds) and the gray curves (lower bounds) are our predictions.}
\label{Fig:emu}
\end{figure}

\begin{figure}[!htbp]
\begin{flushleft}
\includegraphics*[60mm,197mm][300mm,278mm]{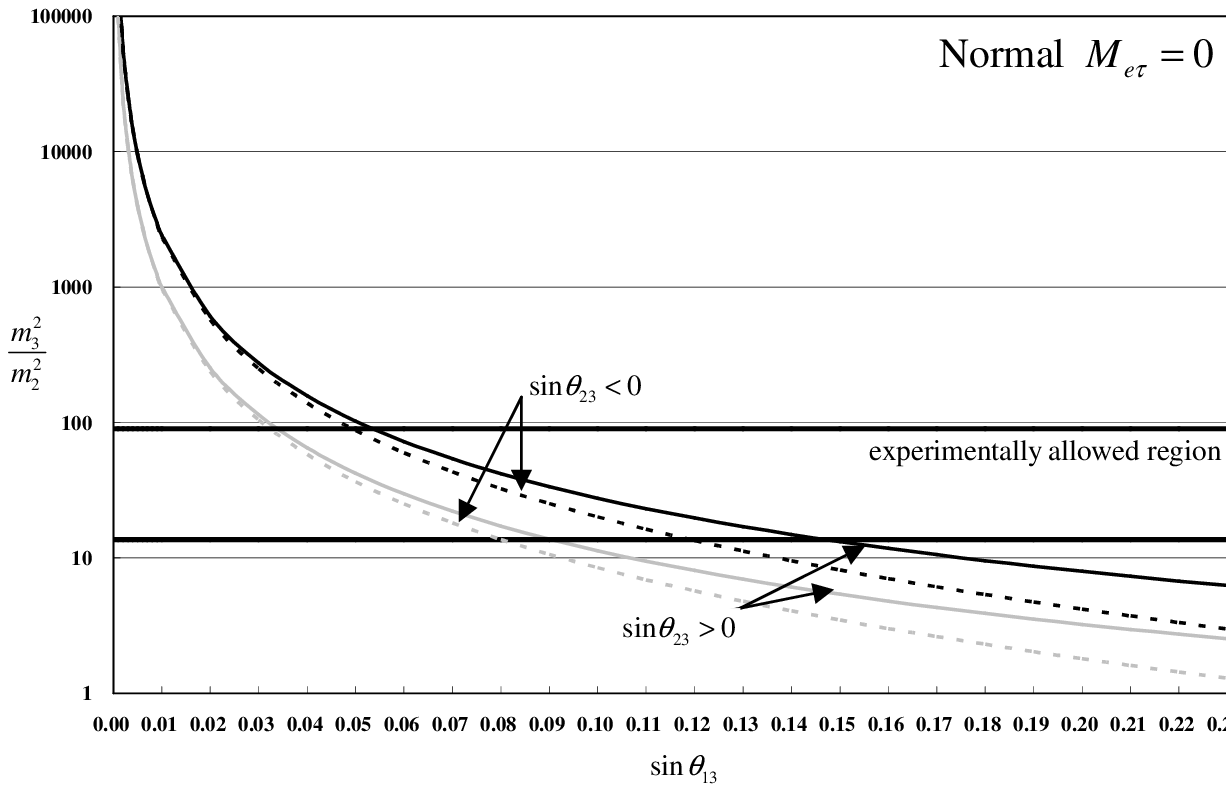}
\end{flushleft}
\caption{The same as in FIG.\ref{Fig:emu} but for $M_{e\tau}=0$.}
\label{Fig:etau}
\end{figure}




\begin{figure}[!htbp]
\begin{flushleft}
\includegraphics*[60mm,197mm][300mm,278mm]{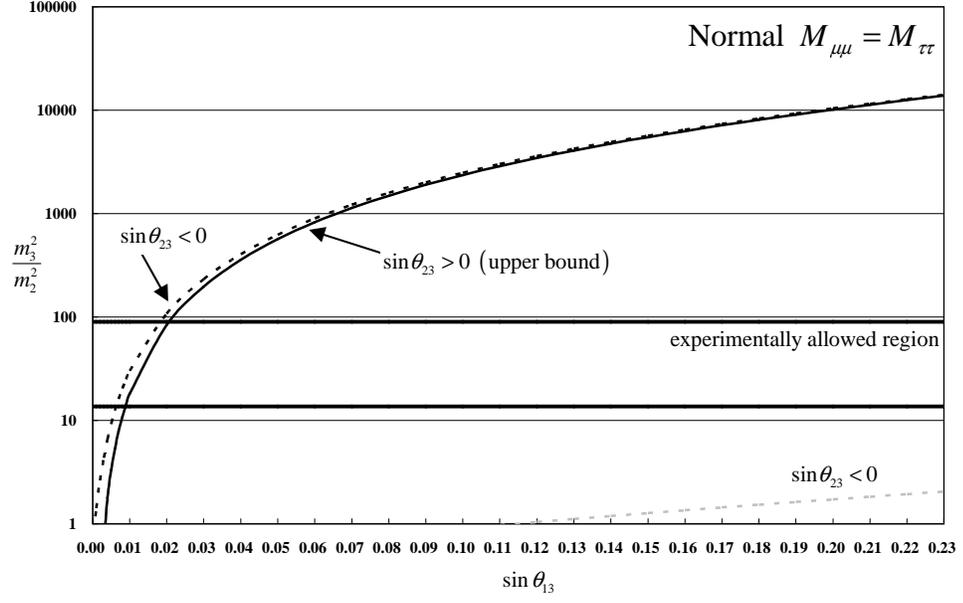}
\end{flushleft}
\caption{The same as in FIG.\ref{Fig:emu} but for $M_{\mu\mu}=M_{\tau\tau}$.  The lower bound in the case of $\sin\theta_{23}>0$ is outside the graph.}
\label{Fig:mumu_tautau}
\end{figure}

\begin{figure}[!htbp]
\begin{flushleft}
\includegraphics*[60mm,197mm][300mm,278mm]{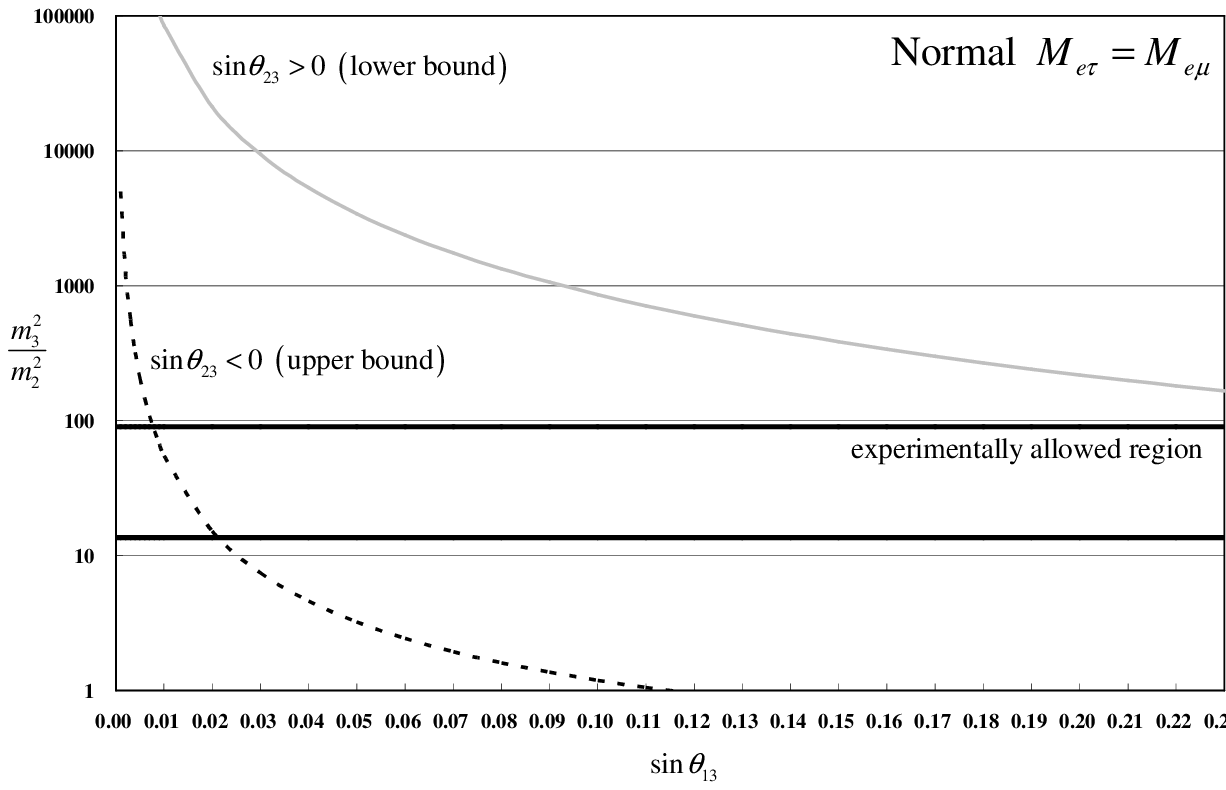}
\end{flushleft}
\caption{The same as in FIG.\ref{Fig:emu} but for $M_{e\tau}=M_{e\mu}$. The allowed region is above the grey solid curve or below the black dashed curve.}
\label{Fig:etau_+emuu}
\end{figure}

\begin{figure}[!htbp]
\begin{flushleft}
\includegraphics*[60mm,197mm][300mm,278mm]{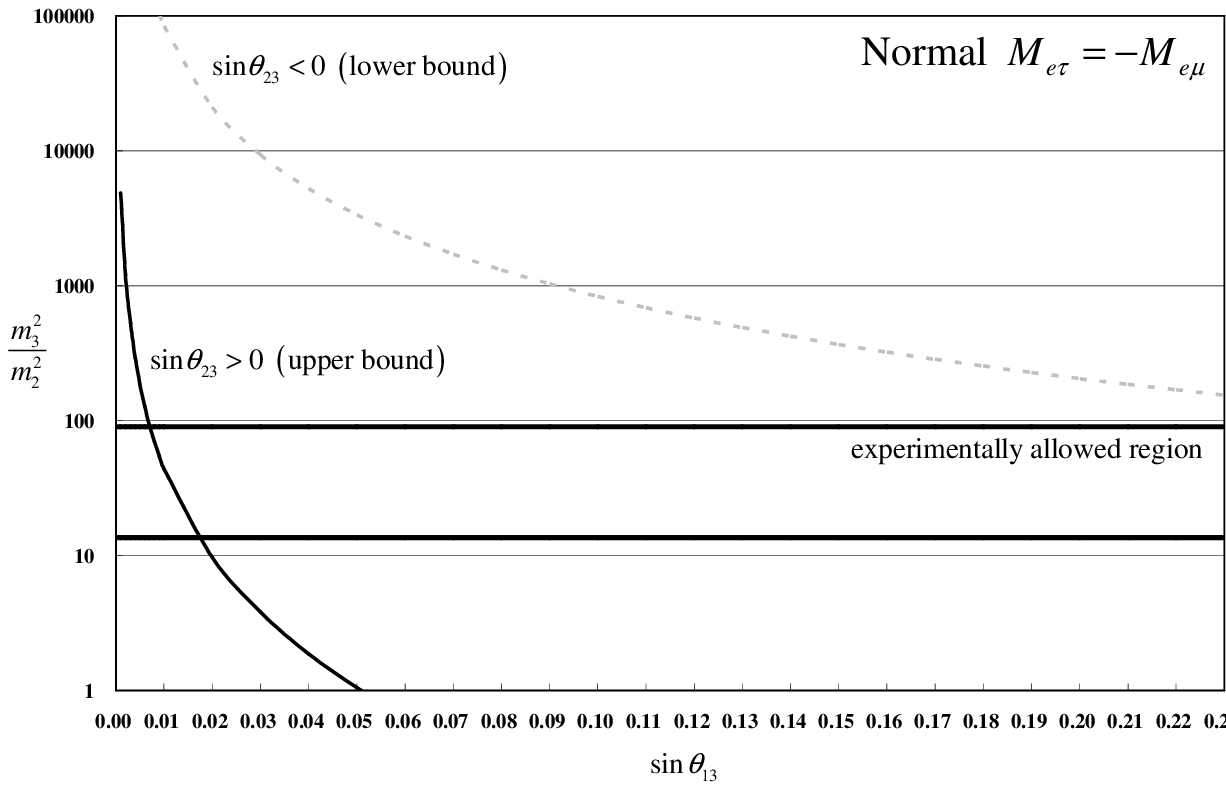}
\end{flushleft}
\caption{The same as in FIG.\ref{Fig:emu} but for $M_{e\tau}=-M_{e\mu}$. The allowed region is above the grey dashed curve or below the black solid curve.}
\label{Fig:etau_-emu}
\end{figure}

\begin{figure}[!htbp]
\begin{flushleft}
\includegraphics*[60mm,197mm][300mm,278mm]{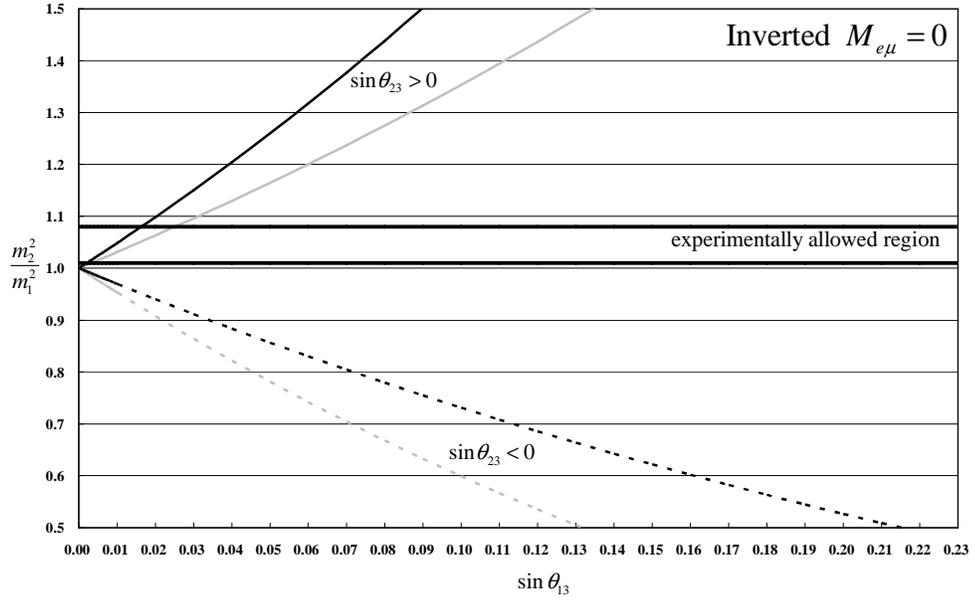}
\end{flushleft}
\caption{The $m^2_2/m^2_1$ for $IMH$: The areas between the black curves (upper bounds) and the gray curves (lower bounds) are our predictions on $m^2_2/m^2_1$ in the texture with $M_{e\mu}=0$.  The solid (dashed) curves correspond to $\sin\theta_{23}>0$ ($\sin\theta_{23}<0$).}
\label{Fig:emu-inv}
\end{figure}

\begin{figure}[!htbp]
\begin{flushleft}
\includegraphics*[60mm,197mm][300mm,278mm]{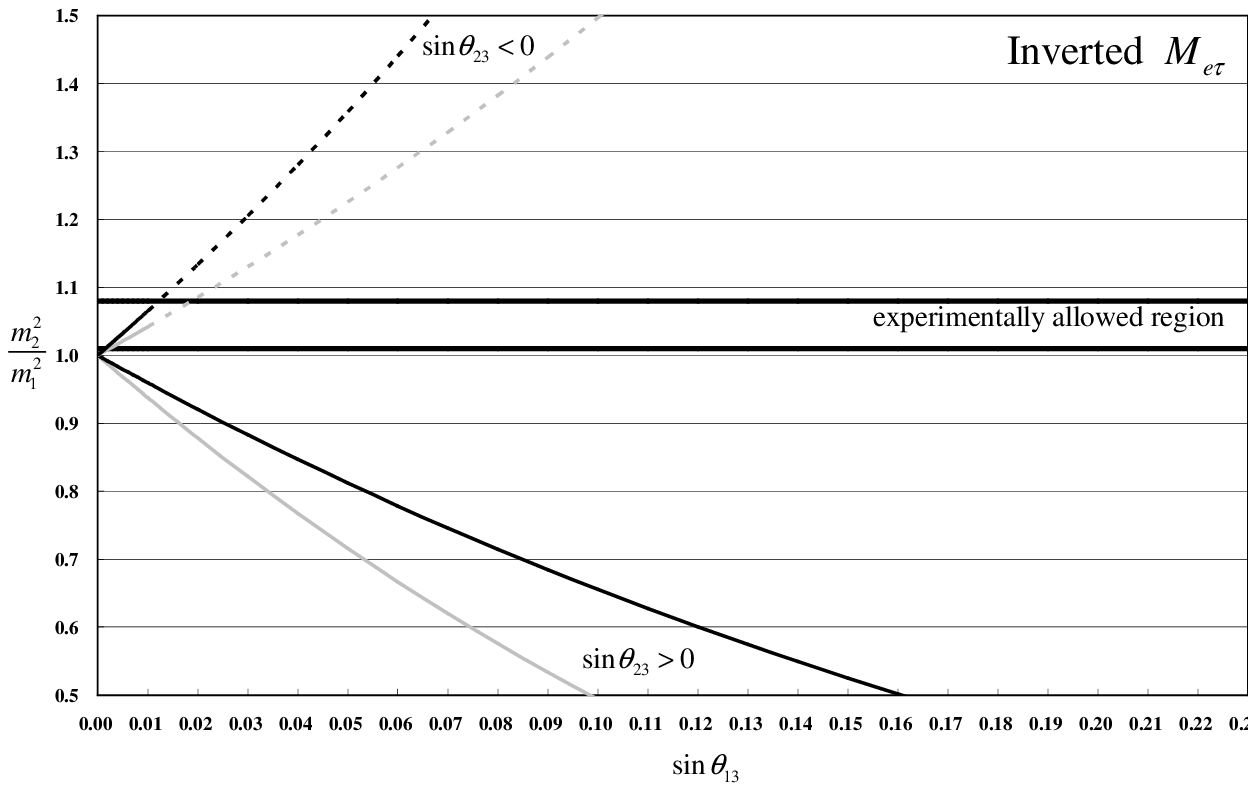}
\end{flushleft}
\caption{The same as in FIG.\ref{Fig:emu-inv} but for $M_{e\tau}=0$.}
\label{Fig:etau-inv}
\end{figure}

\begin{figure}[!htbp]
\begin{flushleft}
\includegraphics*[60mm,197mm][300mm,278mm]{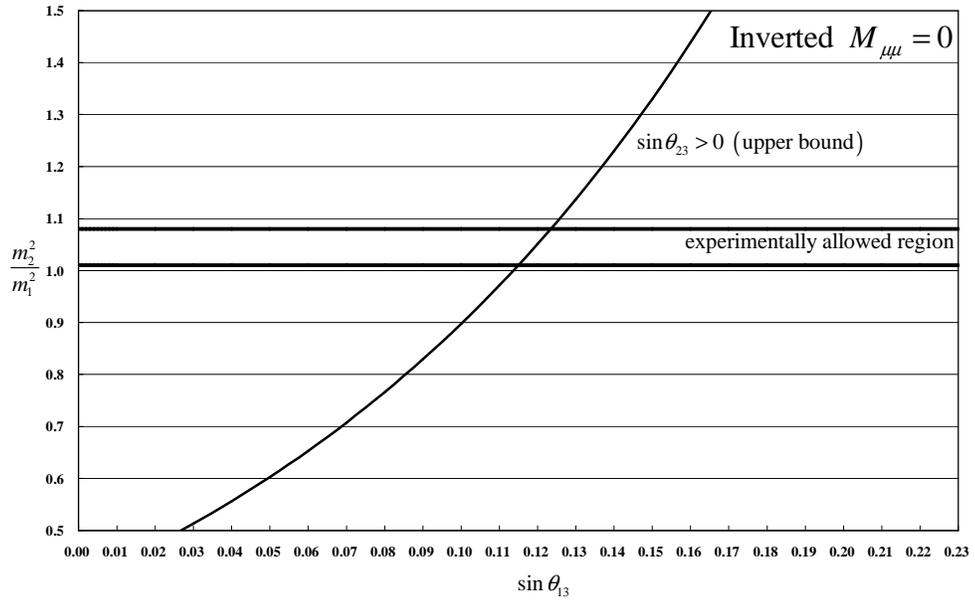}
\end{flushleft}
\caption{The same as in FIG.\ref{Fig:emu-inv} but for $M_{\mu\mu}=0$. The allowed region is below the black solid curve.}
\label{Fig:mumu-inv}
\end{figure}


\begin{figure}[!htbp]
\begin{flushleft}
\includegraphics*[60mm,197mm][300mm,278mm]{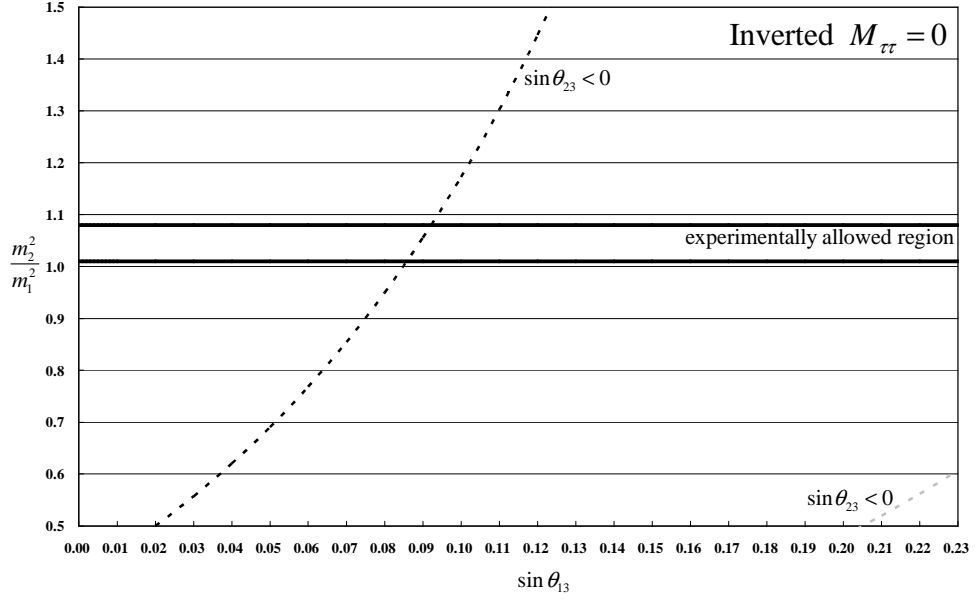}
\end{flushleft}
\caption{The same as in FIG.\ref{Fig:emu-inv} but for $M_{\tau\tau}=0$. The allowed region is below the black dashed curve and above the grey dashed curve.}
\label{Fig:tautau-inv}
\end{figure}

\begin{figure}[!htbp]
\begin{flushleft}
\includegraphics*[60mm,197mm][300mm,278mm]{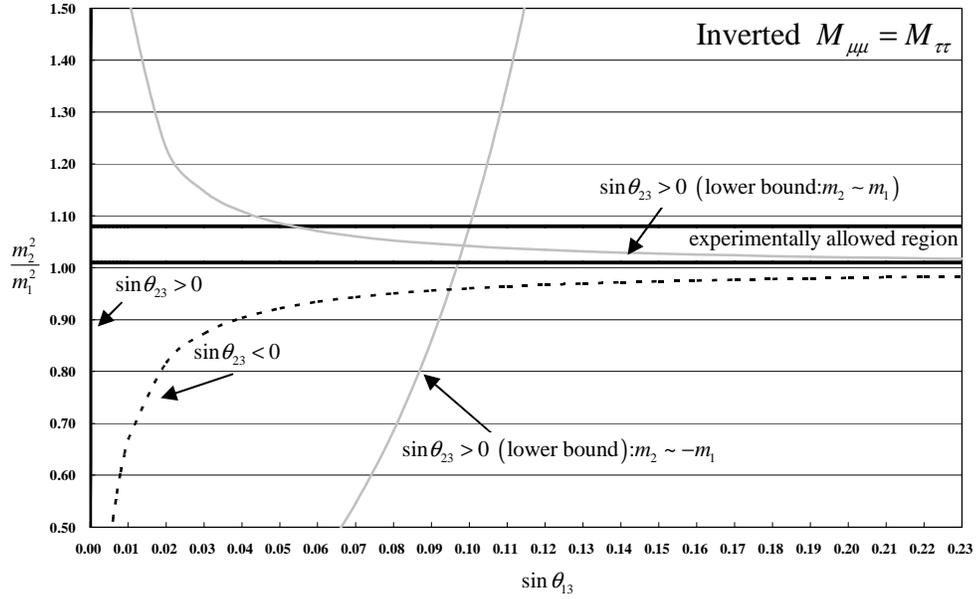}
\end{flushleft}
\caption{The same as in FIG.\ref{Fig:emu-inv} but for $M_{\mu\mu}=M_{\tau\tau}$.  Two grey solid curves represent two lower bounds for the cases with $m_2\sim m_1$ and $m_2\sim -m_1$.  The allowed region is either above one of the grey solid curves that gives the smaller value or below the black dashed curve.  For $\sin\theta_{23}>0$, the upper bound is very steep so that it almost runs on the vertical axis.}
\label{Fig:mumu_tautau-inv}
\end{figure}

\begin{figure}[!htbp]
\begin{flushleft}
\includegraphics*[60mm,197mm][300mm,278mm]{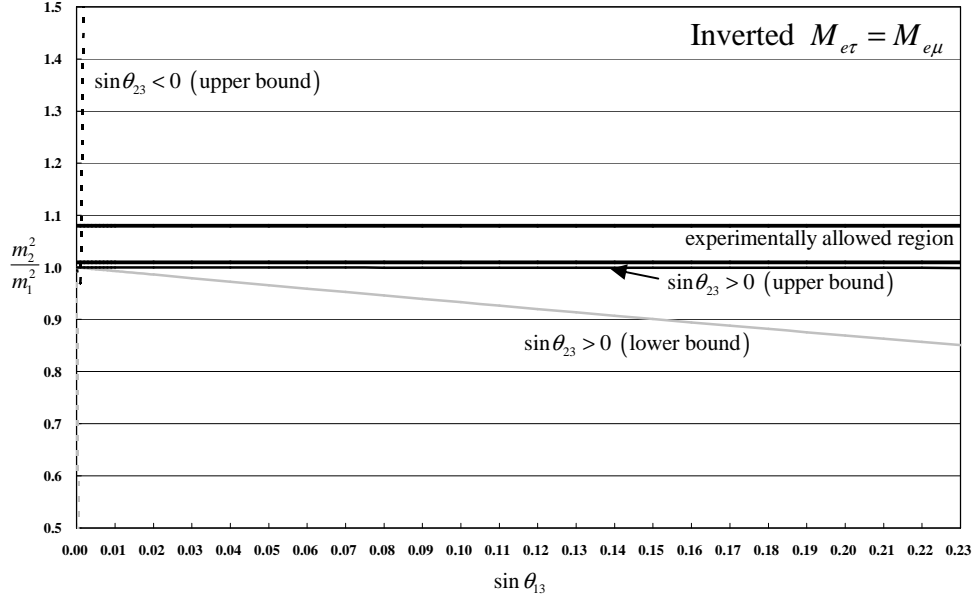}
\end{flushleft}
\caption{The same as in FIG.\ref{Fig:emu-inv} but for $M_{e\tau}=M_{e\mu}$. The allowed regions are below the black dashed curve and the area between the black solid line at $m^2_2/m^2_1=1$ and the gray solid curve. The lower bound in the case of $\sin\vartheta_{23}<0$ is outside the graph.}
\label{Fig:etau_+emuu-inv}
\end{figure}

\begin{figure}[!htbp]
\begin{flushleft}
\includegraphics*[60mm,197mm][300mm,278mm]{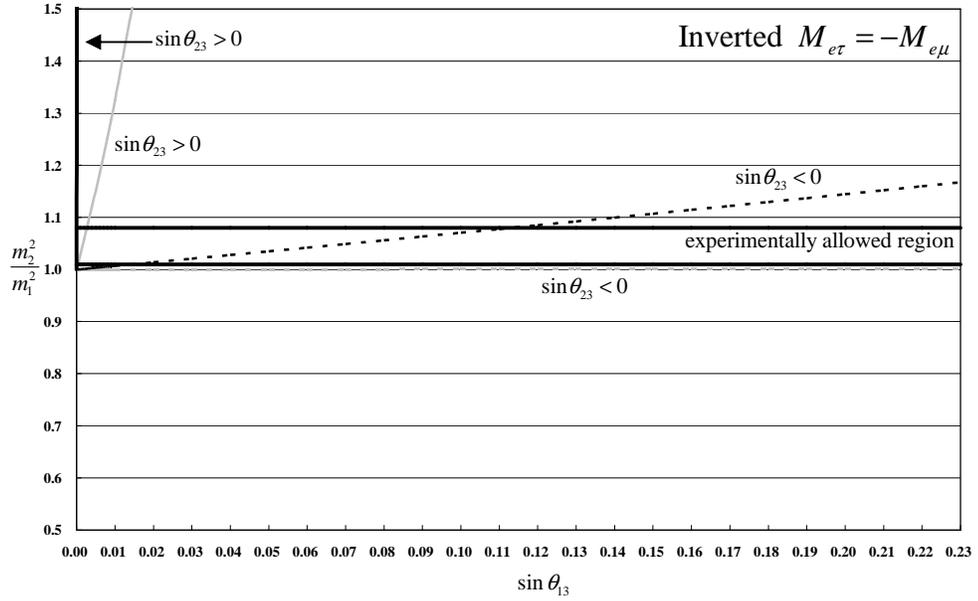}
\end{flushleft}
\caption{The same as in FIG.\ref{Fig:emu-inv} but for $M_{e\tau}=-M_{e\mu}$.  The allowed regions are the area between the black vertical line and the grey curve and the area between the black dashed curve and the gray dashed curve at $m^2_2/m^2_1=1$.  For $\sin\theta_{23}>0$, the upper bound in the case of $\sin\vartheta_{23}>0$ is very steep so that it almost runs on the vertical axis.}
\label{Fig:etau_-emu-inv}
\end{figure}



\begin{thebibliography}{}
\bibitem{SK}
	Y. Fukuda {\it et al.}. [Super-Kamiokande Collaboration], \Journal{\PRL}{81}{1158}{1998};
	[\Journal{\Erratum}{81}{4279}{1998}]; \Journal{\PRL}{82}{2430}{1999}.
	See also,
	T. Kajita and Y. Totsuka, \Journal{\RMP}{73}{85}{2001}.

\bibitem{OldSolar}
    J.N. Bahcall, W.A. Fowler, I. Iben and R.L. Sears, \Journal{\APJ}{137}{344}{1963};
    J. Bahcall, \Journal{\PRL}{12}{300}{1964};
    R. Davis, Jr., \Journal{\PRL}{12}{303}{1964};
    R. Davis, Jr., D.S. Harmer and K.C. Hoffman, \Journal{\PRL}{20}{1205}{1968}; 
    J.N. Bahcall, N.A. Bahcall and G. Shaviv, \Journal{\PRL}{20}{1209}{1968};
    J.N. Bahcall and R. Davis, Jr., \Journal{\SCI}{191}{264}{1976}.

\bibitem{experiments}
	Q.A. Ahmed. {\it et al.}, [SNO Collaboration], \Journal{\PRL}{87}{071301}{2001}; \Journal{\PRL}{89}{011301}{2002};
	S. H. Ahn, {\it et al.}, [K2K Collaboration], \Journal{\PLB}{511}{178}{2001}; \Journal{\PRL}{90}{041801}{2003};
	K. Eguchi, {\it et al.}, [KamLAND collaboration], \Journal{\PRL}{90}{021802}{2003};
	M. Apollonio, {\it et al.}, [CHOOZ Collaboration], \Journal{\EPJ}{27}{331}{2003}.
\bibitem{PMNS} 
	B. Pontecorvo, \Journal{\JETPUSSR}{34}{247}{1958}; \Journal{\ZETP}{53}{1717}{1967};
	Z. Maki, M. Nakagawa and S. Sakata, \Journal{\PTP}{28}{870}{1962}. 

\bibitem{Seesaw} 
	T. Yanagida, in {\it Proceedings of the Workshop on Unified Theories and 
	Baryon Number in the Universe} edited by A. Sawada and A. Sugamoto 
	(KEK Report No.79-18, Tsukuba, 1979), p.95; \Journal{\PTP}{64}{1103}{1980};  
	M. Gell-Mann, P. Ramond and R. Slansky, in {\it Supergravity} edited by P. van Nieuwenhuizen and D.Z. Freedmann (North-Holland, Amsterdam 1979), p.315; 
	R.N. Mohapatra and G. Senjanovi\'{c}, \Journal{\PRL}{44}{912}{1980}. See also,
	P. Minkowski, \Journal{\PLBOLD}{B67}{421}{1977}.

\bibitem{type2seesaw}
	R.N. Mohapatra and G. Senjanovi\'{c}, \Journal{\PRD}{23}{165}{1981};
	C. Wetterich, \Journal{\NPB}{187}{343}{1981}.
	See also, J. Schechter and J.W.F. Valle, \Journal{\PRD}{22}{2227}{1980}.

\bibitem{Zee}
	A. Zee, \Journal{\PLBOLD}{93B}{389}{1980}; \Journal{\PLBOLD}{161B}{141}{1985};
	L. Wolfenstein, \Journal{\NPB}{175}{93}{1980};
	S. T. Petcov, \Journal{\PLBOLD}{115B}{401}{1982}.

\bibitem{Babu}
	A. Zee, \Journal{\NPBOLD}{264B}{99}{1986}; 
	K. S. Babu, \Journal{\PLB}{203}{132}{1988}; 
	D. Chang, W-Y. Keung and P.B. Pal, \Journal{\PRL}{61}{2420}{1988}; 
	J. Schechter and J.W.F. Valle, \Journal{\PLB}{286}{321}{1992}.

\bibitem{NeutrinoSummary}
See for example, 
	R.N. Mohapatra, {\it et al.},``Theory of Neutrinos", [arXive:hep-ph/0412099].
See also,
	S. Goswami, Talk given at {\it Neutrino 2004: The 21st International Conference on Neutrino Physics and Astrophysics}, Paris, France (June 14-19, 2004);
	G. Altarelli, Talk given at {\it Neutrino 2004: The 21st International Conference on Neutrino Physics and Astrophysics}, Paris, France (June 14-19, 2004);
	A. Bandyopadhyay, \Journal{\PLB}{608}{115}{2005}.

\bibitem{PositiveSolor}
O. Mena and S. Parke, \Journal{\PRD}{69}{117301}{2004};

\bibitem{Nishiura}
	T. Fukuyama and H. Nishiura, in {\it Proceedings of International Workshop on Masses and Mixings of Quarks and Leptons} edited by Y. Koide (World Scientific, Singapore, 1997), p.252; ``Mass Matrix of Majorana Neutrinos", [arXive:hep-ph/9702253];
	Y. Koide, H. Nishiura, K. Matsuda, T. Kikuchi and T. Fukuyama, \Journal{\PRD}{66}{093006}{2002};
	Y. Koide, \Journal{\PRD}{69}{093001}{2004};
	K. Matsuda and H. Nishiura, \Journal{\PRD}{69}{117302}{2004}.
\bibitem{mu-tau}
See for example,
	C.S. Lam, \Journal{\PLB}{507}{214}{2001};
	W. Grimus and L. Lavoura,  \Journal{\JHEP}{0107}{045}{2001}; \Journal{\EPJ}{28}{123}{2003}; \Journal{\PLB}{572}{189}{2003}; \Journal{\JPG}{30}{73}{2004};
	T. Kitabayashi and M. Yasu\`{e}, \Journal{\PLB}{524}{308}{2002}; 
	\Journal{\IJMP}{17}{2519}{2002}; \Journal{\PRD}{67}{015006}{2003};
	P.F. Harrison and W.G. Scott, \Journal{\PLB}{547}{219}{2002}; 
	I. Aizawa, M. Ishiguro, T. Kitabayashi and M. Yasu\`{e}, \Journal{\PRD}{70}{015011}{2004};
	W. Grimus, A.S. Joshipura, S. Kaneko, L. Lavoura, H. Sawanaka and M. Tanimoto, \Journal{\JHEP}{0407}{078}{2004};
	``Non-Vanishing $U_{e3}$ and $\cos 2 \theta_{23}$ from a Broken $Z_2$ Symmetry", [arXive:hep-ph/0408123];
\bibitem{mu-tau1}
	R.N. Mohapatra,  \Journal{\JHEP}{0410}{027}{2004};
	R.N. Mohapatra and S. Nasri,  ``Leptogenesis and $\mu-\tau$ symmetry" (to appear in Phys, Rev D), [arXive:hep-ph/0410369];
	R.N. Mohapatra, S. Nasri and H. Yu, ``Leptogenesis, $\mu$-$\tau$ Symmetry and $\vartheta_{13}$", [arXive:hep-ph/0502026].
	

\bibitem{Bimaximal}
	See for example,
	F. Vissani, ``A Study of the Scenario with Nearly Degenerate Majorana Neutrinos", [arXiv:hep-ph/9708483];
	D. V. Ahluwalia, \Journal{\MPL}{13}{2249}{1998};
	V. Barger, P. Pakvasa, T.J. Weiler and K. Whisnant, \Journal{\PLB}{437}{107}{1998};
	A.J. Baltz, A.S. Goldhaber and M. Goldhaber, \Journal{\PRL}{81}{5730}{1998};
	M. Jezabek and Y. Sumino, \Journal{\PLB}{440}{327}{1998};
	R.N. Mohapatra and S. Nussinov, \Journal{\PLB}{441}{299}{1998};, \Journal{\PRD}{60}{013002}{1999};
	Y. Nomura and T. Yanagida, \Journal{\PRD}{59}{017303}{1999};
	I. Starcu and D.V.Ahluwalia, \Journal{\PLB}{460}{431}{1999};
	Q. Shafi and Z. Tavartkiladze, \Journal{\PLB}{451}{129}{1999}; \Journal{\PLB}{482}{145}{2000};
	C.H. Albright and S.M. Barr, \Journal{\PLB}{461}{218}{1999};
	H. Georgi and S.L. Glashow, \Journal{\PRD}{61}{097301}{2000};
	R.N. Mohapatra, A. P\'{e}rez-Lorenzana and C. A. de S. Pires, \Journal{\PLB}{474}{355}{2000};
	T. Kitabayashi and M. Yasu\`{e}, \Journal{\PLB}{490}{236}{2000}; \Journal{\PRD}{63}{095002}{2001};
	\Journal{\PRD}{63}{095006}{2001}; \Journal{\PLB}{508}{85}{2001}; \Journal{\NPB}{609}{61}{2001};
	B. Brahmacari and S. Choubey, \Journal{\PLB}{531}{99}{2002};
	K. S. Babu and R. N. Mohapatra, \Journal{\PLB}{532}{77}{2002};
	T. Ohlsson and G. Seidl, \Journal{\PLB}{537}{95}{2002};
	R. Kuchimanchi and R. N. Mohapatra, \Journal{\PRD}{66}{051301(R)}{2002};
	C. Giunti and M. Tanimoto, \Journal{\PRD}{66}{053013}{2002}.

\bibitem{Lprime}
	R. Barbieri, L.J. Hall, D. Smith, N.J. Weiner and A. Strumia, \Journal{\JHEP}{12}{017}{1998}.
	See also
	S.T. Petcov, \Journal{\PLBOLD}{110B}{245}{1982};
	C.N. Leung and S.T. Petcov, \Journal{\PLBOLD}{125B}{461}{1983};
	A. Zee, \Journal{\PLBOLD}{161B}{141}{1985}.

\bibitem{tri-bimaximal}
See for example,
	P.F. Harrison, D.H. Perkins and W.G. Scott, \Journal{\PLB}{349}{137}{1995}; \Journal{\PLB}{530}{167}{2002};
	Z.-Z. Xing, \Journal{\PLB}{533}{85}{2002};
	P.F. Harrison and W.G. Scott, \Journal{\PLB}{535}{163}{2002}; \Journal{\PLB}{557}{76}{2003}; \Journal{\PLB}{594}{324}{2004};
	C.I. Low and R.R. Volkas, \Journal{\PRD}{68}{033007}{2003}; 
	X.-G. He and A. Zee, \Journal{\PLB}{560}{87}{2003};
	N. Li and Bo-Q. Ma, \Journal{\PRD}{71}{017302}{2005}. 
\bibitem{tri-bimaximal-review}
	P.F. Harrison and W.G. Scott, \Journal{\IJMP}{18}{3957}{2003}.

\bibitem{NearlyBimaximal} 
	H. Fritzsch and Z.Z. Xing, \Journal{\PLB}{372}{265}{1996}; \Journal{\PLB}{440}{313}{1998};
	M. Fukugita, M. Tanimoto and T. Yanagida, \Journal{\PRD}{57}{4429}{1998}; 
	M. Tanimoto, \Journal{\PRD}{59}{017304}{1999}. 

\bibitem{RecentAnalyses}
	M. Apollonio {\it et al.} [CHOOZ Collaboration], \Journal{\EPJ}{27}{331}{2003};
	E. Kearns {\it et al.} [Super-Kamiokande Collaboration], Talk given at {\it Neutrino 2004: The 21st International Conference on Neutrino Physics and Astrophysics}, Paris, France (June 14-19, 2004);
	T. Nakaya {\it et al.} [K2K Collaboration], Talk given at {\it Neutrino 2004: The 21st International Conference on Neutrino Physics and Astrophysics}, Paris, France (June 14-19, 2004);
	T. Araki {\it et al.} [KamLAND Collaboration], ``Measurement of Neutrino Oscillation with KamLAND: Evidence of Spectral Distortion", [arXive:hep-ex/0406035].
\bibitem{ZeroTexture0}
H. Nishiura, K. Matsuda and T. Fukuyama, Phys. Rev. D60, 013006 (1999);
M.-C. Chen and K.T. Mahanthappa, \Journal{\PRD}{62}{113007}{2000};
S.K. Kang and C.S. Kim, \Journal{\PRD}{63}{113010}{2001}.

\bibitem{ZeroTexture}
See for example, 
P.H. Frampton, S.L. Glashow and D. Marfatia, \Journal{\PLB}{536}{79}{2002};
Z. Z. Xing, \Journal{\PLB}{530}{159}{2002}, \Journal{\PLB}{539}{85}{2002}, \Journal{\PRD}{69}{013006}{2004};
P. H. Frampton, S. L. Glashow and T. Yanagida, \Journal{\PLB}{548}{119}{2002};
A. Kageyama, S. Kaneko, N. Shimoyama and M. Tanimoto, \Journal{\PLB}{538}{96}{2002};
M. Raidal and A. Strumia, \Journal{\PLB}{553}{72}{2003};
R. Barbieri, T. Hambye and A. Romanino, \Journal{\JHEP}{0303}{017}{2003};
B.R. Desai, D.P. Roy and A.R. Vaucher, \Journal{\MPL}{18}{1355}{2003};
M. Bando, S. Kaneko, M. Obara and M. Tanimoto, \Journal{\PLB}{580}{229}{2004};
L. Lavoura, \Journal{\PLB}{609}{317}{2005};
W. Grimus and L. Lavoura, ``On a Model with Two Zeros in the Neutrino Mass Matrix", [arXive:hep-ph/0412283].
\bibitem{detM}
	G. C. Branco, R. G. Felipe, F. R. Joaquim and T. Yanagida, \Journal{\PLB}{562}{265}{2003}.
\bibitem{trM}
	D. Black, A. H. Fariborz, S. Nasri and J. Schechter \Journal{\PRD}{62}{073015}{2000};
	X.-G. He and A. Zee, \Journal{\PRD}{68}{037302}{2003};
	W. Rodejohann, \Journal{\PLB}{579}{127}{2004}.
\bibitem{Leptogenesis}
	T. Asaka, M. Fujii, K. Hamaguchi and T. Yanagida, \Journal{\PRD}{62}{123514}{2000};
	M. Fujii, K. Hamaguchi and T. Yanagida, \Journal{\PRD}{64}{123526}{2001}.
\bibitem{Affleck-Dine}
	I. Affleck and M. Dine, \Journal{\NPB}{249}{361}{1985}.
\bibitem{LeptogenesisFirst}
	M. Fukugida and T. Yanagida, \Journal{\PLB}{174}{45}{1986}.
\bibitem{theta-13-th}
See for example,
H. Minakata, H. Sugiyama and O. Yasuda, K. Inoue and F. Suekane, \Journal{\PRD}{68}{033017}{2003};
C. Lunardini and A.Yu. Smirnov, \Journal{\JCAP}{06}{009}{2003};
J. Bernabeu, S.P. Ruiz and S.T. Petcov, \Journal{\NPB}{669}{255}{2003};
O. Yasuda, ``Measurement of $\sin^2{2\theta_{13}}$ by reactor experiments and its sensitivity", Talk presented at Coral Gables Conference On Lauching Of Belle Epoque In High-Energy Physics And Cosmology (CG 2003), 17-21 Dec 2003, Ft. Lauderdale, Florida, [arXive:hep-ph/0403162];
Q. Shafi and Z. Tavartkiladze, \Journal{\PLB}{594}{177}{2004};
O.L.G. Peres and A.Yu. Smirnov, \Journal{\NPB}{680}{479}{2004};
A. Romanino, \Journal{\PRD}{70}{013003}{2004};
J.W. Mei and Z.Z. Xing,  \Journal{\PRD}{70}{053002}{2004};
S. P. Ruiz and S.T. Petcov, \Journal{\NPB}{712}{392}{2005};
N.N. Singh and M.K. Das, ``Radiative Generation of $\Delta m^2_{21}$ and in Two-fold Degenerate Neutrino Models", [arXive:hep-ph/0407206];
J. Ferrandis and S. Pakvasa, \Journal{\PLB}{603}{184}{2004};
S. Goswami and A.Yu. Smirnov. ``Solar Neutrinos and 1-3 Leptonic Mixing", [arXive:hep-ph/0411359].
\bibitem{theta-13-ex}
See for example,
A. Guglielmi, \Journal{\PAN}{67}{1129}{2004};
M. Goodman, ``Plans for Experiments to Measure $\theta_{13}$", [arXive:hep-ex/0404031];
S. Rigolin, ``Why Care about $(\theta_{13},\delta)$ Degeneracy at Future Neutrino Experiments", [arXive:hep-ph/0407009];
Th. Lasserre, ``Chasing Theta(13) with New Reaxtor Neutrino Experiments", [arXive:hep-ex/0411083];
A.Ferrari. A. Guglielmi and P.R, Sala, ``CNGS Neutrino Beam Systematics for $\theta_{13}$, [arXive: hep-ph/0501283].
\bibitem{GeneralM-CP}
I. Aizawa and M. Yasu\`{e}, \Journal{\PLB}{607}{267}{2005}.
\bibitem{AbsoluteMass}
See for example, 
C. Giunti, ``Phenomenology of Absolute Neutrino Masses", talk given at {\it NOW-2004, Neutrino Oscillation Workshop}, Conca Specchiulla, Otranto, Italy (Sep. 11-17, 2004), [arXive: hep-ph/0412148] and references therein.
\bibitem{eNumber}
See for example,
M. Frigerio and A.Yu. Smirnov, \Journal{\NPB}{640}{233}{2002}.
\bibitem{mu-tau2}
	I. Aizawa, M. Ishiguro, T. Kitabayashi and M. Yasu\`{e}, in Ref.\cite{mu-tau}.
\bibitem{ee-exp}
H.V. Klapdor-Kleingrothaus {\it et al.}, \Journal{\EPJA}{12}{147}{2001}.
\end{thebibliography}
\end{document}